\documentclass[a4paper]{article}
\usepackage{amsthm,amsmath,amssymb}
\theoremstyle{plain}
\newtheorem{thm}{Theorem}
\newtheorem{lem}[thm]{Lemma}
\newtheorem{prop}[thm]{Proposition}

\theoremstyle{remark}
\newtheorem*{rem}{Remark}
\newtheorem{exm}{Example}

\usepackage{tikz}
\usetikzlibrary{snakes,shapes}
\tikzstyle{hyb}=[rectangle,fill=green!50,draw,minimum size=2.6mm]
\tikzstyle{hybn}=[rectangle,draw,minimum size=2.6mm]
\tikzstyle{tre}=[circle,fill=green!50,draw,minimum size=3.1mm]
\tikzstyle{treg}=[circle,fill=green!50,draw,minimum size=4mm]
\tikzstyle{trem}=[circle,fill=green!50,draw,minimum size=3.5mm]
\tikzstyle{tres}=[circle,fill=green!50,draw,minimum size=2.5mm]
\tikzstyle{tri}=[rectangle, fill=red!50,draw,minimum size=3mm]
\newcommand{\etq}[1]{%
\draw (#1) node {\scriptsize${#1}$};
}

\newcommand{\pathgr}{\!\rightsquigarrow\!{}}

\renewcommand{\leq}{\leqslant}
\renewcommand{\geq}{\geqslant}

\renewcommand{\le}{\leqslant}
\renewcommand{\ge}{\geqslant}
\newcommand{\NN}{\mathbb{N}}

\DeclareMathOperator{\Tred}{\mathit{T}}
\DeclareMathOperator{\Hred}{\mathit{H}}
\DeclareMathOperator{\TRred}{\mathit{TR}}
\DeclareMathOperator{\HRred}{\mathit{HR}}
\DeclareMathOperator{\indeg}{indeg}
\DeclareMathOperator{\outdeg}{outdeg}

\begin{document}

\title{A Distance Metric for Tree-Sibling Time Consistent Phylogenetic Networks} 

\author{
\textbf{Gabriel Cardona}\\
Department of Mathematics\\
and Computer Science\\
University of the Balearic Islands\\
E-07122 Palma de Mallorca\\
Spain
\and
\textbf{Merc\`e Llabr\'es}\\
Department of Mathematics\\
and Computer Science\\
University of the Balearic Islands\\
E-07122 Palma de Mallorca\\
Spain
\and
\textbf{Francesc Rossell\'o}\\
Department of Mathematics\\
and Computer Science\\
University of the Balearic Islands\\
E-07122 Palma de Mallorca\\
Spain
\and \textbf{Gabriel Valiente}\\
Algorithms, Bioinformatics, Complexity\\
and Formal Methods Research Group\\
Technical University of Catalonia\\
E-08034 Barcelona\\
Spain
}


\maketitle

\begin{abstract}

\noindent{}\textbf{Motivation:}
The presence of reticulate evolutionary events in phylogenies turn
phylogenetic trees into phylogenetic networks. 
These events imply in particular that there may exist multiple
evolutionary paths from a non-extant species to an extant one, and
this multiplicity makes
the comparison of phylogenetic networks much more difficult than
the comparison of phylogenetic trees.
In fact, all attempts
to define a sound distance measure on the 
class of all phylogenetic
networks have failed so far.
Thus, the only practical solutions have been either
the use of rough estimates of similarity (based on comparison of the
trees embedded in the networks), or narrowing the class of phylogenetic
networks to a certain class where such a distance is known and can be
efficiently computed. The first approach has the problem that one may
identify two networks as equivalent, when they are not; the second one
has the drawback that there may not exist algorithms to reconstruct such
networks from biological sequences.

\noindent{}\textbf{Results:}
We present in this paper a distance measure on the class of
\emph{tree-sibling time consistent}
phylogenetic networks, which generalize tree-child
time consistent phylogenetic networks, and thus also galled-trees.
The practical
interest of this distance measure is twofold: it can be computed in
polynomial time by means of simple algorithms, and there also exist
polynomial-time algorithms for reconstructing networks of this class
from DNA sequence data.

\noindent{}\textbf{Availability:}
The Perl package \texttt{Bio::PhyloNetwork}, included in the BioPerl
bundle,  implements many algorithms
on phylogenetic networks, including the computation
of the distance presented in this paper.

\noindent{}\textbf{Contact:}
\texttt{gabriel.cardona@uib.es}
\end{abstract}

\section{Introduction}

Phylogenies reveal the history of evolutionary events of a group of
species, and they are central to comparative analysis methods for
testing hypotheses in evolutionary
biology~\cite{pagel:1999}. Although phylogenetic trees have been used
since the early days of phylogenetics~\cite{darwin:1837} to represent
evolutionary histories under mutation,
it is currently well known that the existance of genetic
recombinations, hybridizations and lateral gene transfers makes
species evolve more in a reticulate way that in a simple, arborescent
way~\cite{doolittle:99}.

Now, as it happens in the case of phylogenetic trees, given a set of
operational taxonomic units, different reconstruction algorithms, or
different sets of sampled data, may lead to different
reticulate evolutionary
histories. Thus, a well-defined distance measure for phylogenetic
networks becomes necessary.  

In a completely general setting, a phylogenetic network is simply a
directed acyclic graph whose leaves (nodes without outgoing edges) are
labeled by the species they
represent~\cite{strimmer.ea:2000,strimmer.ea:2001}. However, this
situation is so general that even the problem of deciding when two
such graphs are isomorphic is computationally hard. Hence, one has to
put additional constraints to narrow down the class of phylogenetic
networks. There have been different approaches to this problem in the
literature, giving rise to different definitions of phylogenetic
network; see 
\cite{bandelt:94,huson:tutgcb06,huson:07,nakhleh.ea:tutpsb04,
  semple:07,strimmer.ea:2000,strimmer.ea:2001}. 

In this paper, we give a distance measure on the class of
\emph{tree-sibling time consistent} phylogenetic networks. This class
first appeared
in Nakhleh's thesis 
\cite{nakhleh:phd04}, and it is of special interest because
there exist algorithms to reconstruct phylogenetic networks of this
class from the analysis of biological sequences
\cite{nakhleh.ea:bioinfo06,nakhleh.ea:bioinfo07}.
However, all previous attempts to provide a sound distance measure
on this class of networks have failed \cite{cardona.ea:math.biosci:2008}.

\section{Tree-sibling time consistent phylogenetic networks}
\label{sec:tree-sibling-phn}

Let $N=(V,E)$ be a
directed acyclic graph, or DAG for short.
We will say that a node $u$ is a
\emph{tree node} if $\indeg(u)\le 1$; moreover, if $\indeg(u)=0$, we will
say that $u$ is a \emph{root} of $N$. If a single root exists, we will
say that the DAG is \emph{rooted}.
We will say that a node $u$ is a 
\emph{hybrid node} if $\indeg(u)\ge 2$. A node $u$ is a
\emph{leaf} if $\outdeg(u)=0$.

In a DAG $N=(V,E)$, we will say that $v$ is a \emph{child} of $u$
if $(u,v)\in E$; in this case, we will also say
that $u$ is a \emph{parent} of $v$.
Note that
any tree node has a single parent, except for the roots of the graph.

Whenever there exists a directed path (eventually trivial) from a node
$u$ to $v$, we will say that $v$ is a \emph{descendant} of $u$, or
that $u$ is an \emph{ancestor} of $v$.

We will say that two nodes $u$ and $v$ are \emph{siblings} of each
other if they share a parent.
Note
that the relation of being siblings is reflexive and symmetric, but
not transitive.

\begin{figure}
  \centering
    \begin{tikzpicture}[thick,>=stealth,scale=0.5]
    \draw(0,0) node[tre] (x) {};
    \draw(0,0) node[hybn] (x) {};
    \draw(-1,-2) node[tre] (u) {};\etq u
    \draw(1,-1) node[hyb] (y) {};
    \draw(1,-2) node[tre] (v) {};\etq v
    \draw[->](x)--(u);
    \draw[->](x)--(y);
    \draw[->](y)--(v);
  \end{tikzpicture}
 \caption{\label{fig:quasi-siblings}Node $v$ is quasi-sibling of $u$.}
\end{figure}
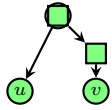

We will say that a tree node $v$ is \emph{quasi-sibling} of another
tree node $u$ if the parent of $v$ is a hybrid node that is also a
sibling of $u$: see Fig.~\ref{fig:quasi-siblings}\footnote{Henceforth,
  in graphical representations of phylogenetic networks, 
hybrid nodes are represented by squares, tree nodes by
circles, and indeterminate nodes (that is, that can be either
tree or hybrid nodes) by both of them superposed.}.
The relation of being quasi-siblings is
neither reflexive nor symmetric.

A \emph{phylogenetic network} on a set $S$ of labels is a rooted DAG
such that: 
\begin{itemize}
\item No tree node has out-degree $1$.
\item Every hybrid node has out-degree $1$, and its single child is a
  tree node.
\item Its leaves are bijectively labeled by $S$.
\end{itemize}
Moreover, if all hybrid nodes have in-degree equal to two, we will say
that
it is a \emph{semi-binary phylogenetic network}. Note that
semi-binarity does not impose
any further condition on the out-degree of tree nodes.

The underlying motivation for such definitions is that tree nodes
represent species, the leaves corresponding to extant ones, and
the internal tree nodes to 
ancestral ones. Hybrid nodes model
recombination events,
where the parents of a hybrid node correspond to the species involved
in this process,
and its single child corresponds to the
resulting species. Hence, the semi-binarity condition means that
these events always involve two, and only two, species.

Although in real applications of phylogenetic networks, the set $S$
labeling the leaves would correspond to a given set of taxa of
extant species,
for the sake of simplicity we will hereafter assume that the set of
labels is simply $S=\{1,\dots,n\}$.

We will say that a phylogenetic network is \emph{tree-sibling} if each
hybrid node has at least one sibling that is a tree node.

Biologically, this condition means that
for each of the hybridization processes,
at least one of the species
involved in it has also some
descendant through 
mutation. 


A \emph{time assignment} on a network $N=(V,E)$ is a mapping
$\tau:V\to\NN$ such that:
\begin{enumerate}
\item $\tau(r)=0$, where $r$ is the root of $N$.
\item If $v$ is a hybrid node and $(u,v)\in E$, then $\tau(u)=\tau(v)$.
\item If $v$ is a tree node and $(u,v)\in E$, then $\tau(u)<\tau(v)$.
\end{enumerate}
We will say that a network is \emph{time consistent} if it admits a
time assignment \cite{baroni.ea:sb06}.

From a biological point of view, a time assignment represents
the time when 
a certain species exists, or a certain hybridization process occurs.
Note that whenever such a process takes place, the species involved
must coexist; this is what the time-consistency property ensures.


By a \emph{sbTSTC network} we will mean a semi-binary
tree-sibling, time consistent phylogenetic network, and this will be
the class of 
phylogenetic networks that we will consider in the rest of the paper. 

\begin{rem}
  Besides the biological considerations we have made while
  presenting our assumptions on phylogenetic networks,
  these are also motivated by the
  fact that we want to single out phylogenetic networks by means of
  their $\mu$-representation (see section~\ref{sec:mu-representation}
  below). In section~\ref{sec:counterexamples} we give examples showing
  that the technical conditions imposed on phylogenetic networks are
  necessary to achieve this goal.
\end{rem}

\begin{figure}
  \centering
  \begin{tikzpicture}[thick,>=stealth,scale=0.6]
    \draw(0,0) node[tre] (r) {}; \etq r;
    \draw(-2,-1) node[tre] (u) {}; \etq u;
    \draw(0,-1) node[tre] (v) {}; \etq v;
    \draw(2,-1) node[tre] (w) {}; \etq w;
    \draw(-1,-1) node[hyb] (A) {}; \etq A;
    \draw(1,-1) node[hyb] (B) {}; \etq B;
    \draw(-2,-2) node[tre] (1) {}; \etq 1;
    \draw(-1,-2) node[tre] (2) {}; \etq 2;
    \draw(1,-2) node[tre] (3) {}; \etq 3;
    \draw(2,-2) node[tre] (4) {}; \etq 4;
    \draw[->] (r)--(u);
    \draw[->] (r)--(v);
    \draw[->] (r)--(w);
    \draw[->] (u)--(A);
    \draw[->] (v)--(A);
    \draw[->] (v)--(B);
    \draw[->] (w)--(B);
    \draw[->] (u)--(1);
    \draw[->] (A)--(2);
    \draw[->] (B)--(3);
    \draw[->] (w)--(4);
  \end{tikzpicture}
  \caption{\label{fig:exm-TSTC}A sbTSTC phylogenetic network.}
\end{figure}
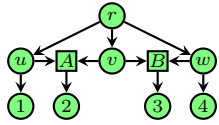

\begin{rem}
We have mentioned in the introduction that the class of semi-binary
tree-sibling time consistent phylogenetic networks generalizes those
introduced in \cite{nakhleh:phd04}. Namely, the latter are obtained
from a phylogenetic tree
by repeating the following procedure:
\begin{enumerate}
\item choose a pair of arcs $(u_1,v_1)$ and $(u_2,v_2)$ in the tree;
\item split these arcs by introducing intermediate nodes $w_1$ (that
  will become a tree node) and $w_2$ (that will become a hybrid node),
  respectively;
\item add a new arc $(w_1,w_2)$.
\end{enumerate}
Each hybrid node introduced, $w_2$ in the notations above, has a tree
sibling, namely $v_1$.
Hence, the networks obtained by this procedure are sbTSTC
networks. However, the sbTSTC network $N_3$ in
Fig.~\ref{fig:Nn}
cannot be obtained by the procedure above
from a tree $T$. 
Indeed, the described procedure cannot introduce tree nodes with
out-degree greater that $2$; hence node $a$ in $N_3$
should also be a node of $T$, and
the out-degree of $r$ in
$T$ would be $1$, yielding to a contradiction.
\end{rem}

The following result ensures the
existence of sibling or quasi-sibling leaves in sbTSTC networks.

\begin{lem}\label{thm:existance}
  Let $N$ be a sbTSTC network. Then, there exists at least one pair of
  leaves 
  that are either siblings or quasi-siblings.
\end{lem}

\begin{proof}
  Let $M$ be the set of internal nodes of $N$ with maximal time
  assignment.

  If no node of $M$ is hybrid, let $u\in M$ be any
  tree node. Then, all its children are leaves: indeed, if a child of
  $u$ were an internal tree node, then its time assignment would
  be strictly greater than that of $u$, against our assumption; also,
  if a child of $u$ were a hybrid node, then its time assignment
  would be the same as that of $u$, and hence $M$ would contain a
  hybrid node. Therefore, since we do not allow out-degree 1 tree
  nodes, the node $u$ has at least two children that are leaves, and
  these leaves are siblings.

  If $M$ contains a hybrid node $v$, then its parents are tree nodes
  $u,u'$ with the same time assignment as that of $v$, and at least one of
  them must have a tree child because 
  of the tree-sibling property.
  Say that $u$ has a tree child; the
  same argument as before proves that this child must be a
  leaf $i$. Moreover, the single child of $v$ must be a tree node,
  hence also a leaf $j$. In this situation we have that $j$ is a
  quasi-sibling of $i$. 
\end{proof}

We give now tight bounds for the number of hybrid and
internal tree nodes of a sbTSTC phylogenetic network, depending on its
number of leaves. The existance of such bounds implies, in
particular, that there exists a finite number of sbTSTC phylogenetic
networks on a given set of taxa up to isomorphisms.
Nevertheless, we
have not yet been able to find a closed expression for this number of
networks depending only on the number of leaves.
Table~\ref{tab:number-networks} shows the experimental results we have
found in this direction
using the procedure described in  Section~\ref{sec:comp-aspects}.

\begin{table}
  \centering
  \caption{\label{tab:number-networks}Number of sbTSTC networks for
    small number $n$ of leaves.}
  \begin{tabular}{r||c|c|c|c|c|}
    $n$  & 1 & 2 & 3 & 4 & 5 \\ \hline
    Number of networks & 1 & 1 & 10 & 606 & 215\,283\\ 
  \end{tabular}
\end{table}

\begin{prop}
  Let $N$ be a sbTSTC network. Let $n,h,t$ be, respectively, the
  number of leaves, the number of hybrid nodes and the number of
  internal tree nodes of $N$.
  If $n\le 2$, then $h=0$ and $t=n-1$. Otherwise,
  $h\le 2n-4$ and $t\le 3n-6$.
\end{prop}

\begin{proof}
  The result is obvious if $n\le 2$, since then $N$ is a tree.

    Assume that $n\ge 3$ and that the result is proved for networks with
  less than $n$ leaves. Let $M$ be the set of internal nodes with
  maximum time 
  assignment, and let $M_t$ (respectively, $M_h$) be the set of tree
  nodes (respectively, hybrid nodes) in $M$.
  Notice that $M_t$ is non-empty,
  because if a hybrid node has maximum time
  assignment, its two parents have the same time assignment and,
  therefore, are in $M_t$. Consider the following different situations:
  \begin{enumerate}
  \item If some node $u$ in $M_t$ has two (or more) children leaves, let
    $N'$ be the sbTSTC network obtained by removing one of these
    leaves and eventually collapsing the created elementary path into
    a single arc. Then the number of leaves, hybrid nodes and internal
    tree nodes in $N'$ is
    $$n'=n-1,\qquad h'=h,\qquad t'=t-\epsilon,$$
    with $\epsilon=0$ if the out-degree of $u$ in $N$ is greater than
    two, and $\epsilon=1$ otherwise. Now, from the induction
    hypothesis we get
    $$
    \begin{aligned}
      h&=h'\le 2n'-4=2n-2-4<2n-4,\\
      t&=t'+\epsilon\le 3n'-6+\epsilon=3n-9+\epsilon<3n-6.
    \end{aligned}
    $$
  \item If (1) does not hold, but every node in $M_t$ has one child
    leaf, let $N'$ be the sbTSTC network obtained by removing all the
    nodes in $M_h$, together with their respective children leaves (say
    $k=|M_h|$), and collapsing the created elementary paths into
    single arcs. In this case we have that
    $$n'=n-k,\qquad h'=h-k,\qquad t'=t-\tilde k,$$
    where $\tilde k\le 2k$ is the number of elementary paths that have
    been removed. Now, also from the induction hypothesis we get
    $$
    \begin{aligned}
      h&=h'+k\le 2n'-4+k=
      2n-2k-4+k=2n-4-k< 2n-4,\\
      t&=t'+\tilde k \le 3n'-6+\tilde k =
      3n-3k-6+\tilde k<3n-6.
    \end{aligned}
    $$
  \item If neither (1) nor (2) hold, then there exists a node $u\in
    M_t$ such that all its children, say $v_1,\dots,v_k$ ($k\ge 2$),
    are in 
    $M_h$. Let $N'$ be the sbTSTC network obtained by removing all
    nodes $v_1,\dots,v_k$ together with their respective children
    leaves, and collapsing the created elementary paths into
    single arcs.
    Notice that the node $u$ is no longer an internal tree node, but a
    leaf of $N'$.
    Then, the number of nodes in $N'$ is
    $$n'=n-k+1,\qquad h'=h-k,\qquad t'=t-\tilde k-1,$$
    where $\tilde k\le k$ is the number of elementary paths that have
    been removed.
    Now, the induction hypothesis yields
    $$
    \begin{aligned}
      h&=h'+k\le 2n'-4+k=2n-2k+2-4+k=
      2n-k-2\le 2n-4,\\
      t&=t'+\tilde k+1\le 3n'-6+\tilde k+1=
      3n-2-3k+\tilde k \le 3n-2-2k
      \le 3n-6.
    \end{aligned}
    $$
\end{enumerate}
  Hence, in all cases, the result follows.
  
\end{proof}

The bounds in the proposition above are tight, as the following
example shows. 

\begin{exm}\label{ex:maximum}
Consider the family of sbTSTC phylogenetic networks $(N_n)_{n\geq 3}$
defined recursively in the following way: 
\begin{itemize}
\item $N_3$ is the first phylogenetic network depicted in Fig.~\ref{fig:Nn}.
\item The network $N_{n+1}$ is obtained from $N_n$ by applying the
  transformation described in Fig.~\ref{fig:op}.
  Fig.~\ref{fig:Nn} depicts also
  $N_4$ and $N_5$, where we
  label the 
  internal nodes in these networks to ease understanding of the
  construction.
\end{itemize}
Note that all networks $N_n$ are semi-binary and tree-sibling by
construction. Also, the time consistency property can be easily
verified: when constructing $N_{n+1}$ from $N_n$, we can assign to
each of the internal nodes introduced the maximum of the times that
the leaves $1,2,n$ have in $N_n$, and reassign to the leaves
$1,2,n,n+1$ this maximum plus one.

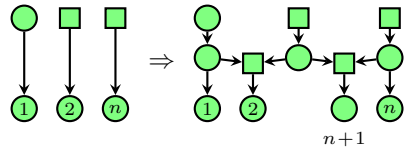
\begin{figure}[htb]
\centering 
  \begin{tikzpicture}[thick,>=stealth,scale=0.3,baseline=0pt]
    \draw (0,0) node[tre] (1) {}; \etq{1}
    \draw (2,0) node[tre] (2) {}; \etq{2}
     \draw (4,0) node[tre] (n) {}; \etq{n}
    \draw (0,4) node[tre] (a) {};  
    \draw (2,4) node[hyb] (A) {};  
     \draw (4,4) node[hyb] (B) {}; 
     \draw [->](a)--(1);
     \draw [->](A)--(2);
     \draw [->](B)--(n);
  \end{tikzpicture}
     \begin{tikzpicture}[thick,>=stealth,scale=0.3,baseline=0pt]
    \draw (0,0) node {};
    \draw (0,2) node {$\Rightarrow$};
    \draw (0,4) node {};
  \end{tikzpicture}
  \begin{tikzpicture}[thick,>=stealth,scale=0.3,baseline=0pt]
    \draw (0,0) node[tre] (1) {}; \etq{1}
    \draw (2,0) node[tre] (2) {}; \etq{2}
     \draw (6,0) node[tre,label=below:{\scriptsize${{n\!+\!1}}$}] (n+1) {}; 
     \draw (8,0) node[tre] (n) {}; \etq{n}
   \draw (0,4) node[tre] (a) {};  
    \draw (4,4) node[hyb] (A) {};  
     \draw (8,4) node[hyb] (B) {}; 
   \draw (0,2.25) node[tre] (b) {};  
    \draw (2,2) node[hyb] (C) {};  
     \draw (6,2) node[hyb] (D) {}; 
   \draw (4,2.25) node[tre] (c) {};  
   \draw (8,2.25) node[tre] (d) {};  
     \draw [->](a)--(b);
     \draw [->](b)--(1);
   \draw [->](B)--(d);
     \draw [->](d)--(n);
    \draw [->](A)--(c);
     \draw [->](c)--(C);
      \draw [->](c)--(D);
     \draw [->](b)--(C);
      \draw [->](d)--(D);
     \draw [->](C)--(2);
      \draw [->](D)--(n+1);
  \end{tikzpicture}
    
\caption{\label{fig:op}  The transformation that produces $N_{n+1}$ from $N_n$.}
\end{figure}

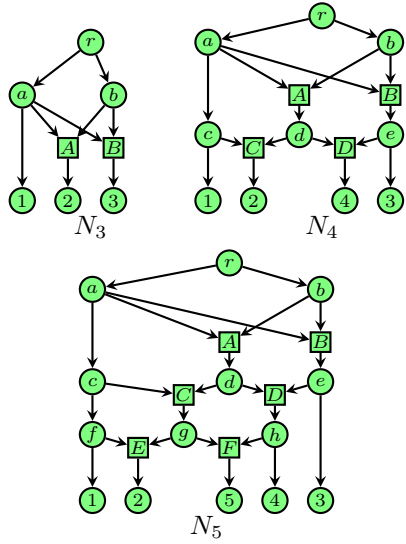
\begin{figure}[htb]
\centering 
  \begin{tikzpicture}[thick,>=stealth,xscale=0.3,yscale=0.35]
    \draw (0,0) node[tre] (1) {}; \etq{1}
    \draw (2,0) node[tre] (2) {}; \etq{2}
     \draw (4,0) node[tre] (3) {}; \etq{3}
    \draw (2,2) node[hyb] (A) {};  \etq{A}
     \draw (4,2) node[hyb] (B) {}; \etq{B}
    \draw (0,4) node[tre] (a) {}; \etq{a}
         \draw (4,4) node[tre] (b) {};  \etq{b}
             \draw (3,6) node[tre] (r) {}; \etq{r}
     \draw [->](r)--(a);
 \draw [->](r)--(b);
     \draw [->](a)--(1);
     \draw [->](a)--(A);
     \draw [->](a)--(B);
    \draw [->](b)--(A);
     \draw [->](b)--(B);
    \draw [->](A)--(2);
     \draw [->](B)--(3);
\draw(3,-1) node {$N_3$};
  \end{tikzpicture}
\qquad
\begin{tikzpicture}[thick,>=stealth,xscale=0.3,yscale=0.35]
    \draw (0,0) node[tre] (1) {}; \etq{1}
    \draw (2,0) node[tre] (2) {}; \etq{2}
      \draw (6,0) node[tre] (4) {}; \etq{4}
    \draw (8,0) node[tre] (3) {}; \etq{3}
    \draw (2,2) node[hyb] (C) {};  \etq{C}
     \draw (6,2) node[hyb] (D) {}; \etq{D}
    \draw (0,2.5) node[tre] (c) {}; \etq{c}
         \draw (4,2.5) node[tre] (d) {};  \etq{d}
             \draw (8,2.5) node[tre] (e) {}; \etq{e}
    \draw (4,4) node[hyb] (A) {};  \etq{A}
     \draw (8,4) node[hyb] (B) {}; \etq{B}
    \draw (0,6) node[tre] (a) {};  \etq{a}
     \draw (8,6) node[tre] (b) {}; \etq{b}
    \draw (5,7) node[tre] (r) {};  \etq{r}
       \draw (5,-1) node {$N_4$}; 
     \draw [->](r)--(a);
 \draw [->](r)--(b);
    \draw [->](a)--(A);
 \draw [->](a)--(c);
    \draw [->](a)--(B);
    \draw [->](b)--(A);
    \draw [->](b)--(B);
    \draw [->](A)--(d);
    \draw [->](B)--(e);
    \draw [->](c)--(1);
    \draw [->](c)--(C);
    \draw [->](d)--(C);
    \draw [->](d)--(D);
    \draw [->](e)--(D);
    \draw [->](e)--(3);
    \draw [->](C)--(2);
    \draw [->](D)--(4);
  \end{tikzpicture}
\\  \begin{tikzpicture}[thick,>=stealth,xscale=0.3,yscale=0.35]
    \draw (0,-2) node[tre] (1) {}; \etq{1}
    \draw (2,-2) node[tre] (2) {}; \etq{2}
      \draw (6,-2) node[tre] (5) {}; \etq{5}
    \draw (8,-2) node[tre] (4) {}; \etq{4}
     \draw (10,-2) node[tre] (3) {}; \etq{3}
    \draw (2,0) node[hyb] (E) {};  \etq{E}
     \draw (6,0) node[hyb] (F) {}; \etq{F}
   \draw (0,0.5) node[tre] (f) {}; \etq{f}
         \draw (4,0.5) node[tre] (g) {};  \etq{g}
             \draw (8,0.5) node[tre] (h) {}; \etq{h}
    \draw (4,2) node[hyb] (C) {};  \etq{C}
     \draw (8,2) node[hyb] (D) {}; \etq{D}
    \draw (0,2.5) node[tre] (c) {}; \etq{c}
         \draw (6,2.5) node[tre] (d) {};  \etq{d}
             \draw (10,2.5) node[tre] (e) {}; \etq{e}
    \draw (6,4) node[hyb] (A) {};  \etq{A}
     \draw (10,4) node[hyb] (B) {}; \etq{B}
    \draw (0,6) node[tre] (a) {};  \etq{a}
     \draw (10,6) node[tre] (b) {}; \etq{b}
    \draw (6,7) node[tre] (r) {};  \etq{r}
       \draw (5,-3) node {$N_5$}; 
     \draw [->](r)--(a);
 \draw [->](r)--(b);
    \draw [->](a)--(A);
 \draw [->](a)--(c);
    \draw [->](a)--(B);
    \draw [->](b)--(A);
    \draw [->](b)--(B);
    \draw [->](A)--(d);
    \draw [->](B)--(e);
    \draw [->](c)--(f);
     \draw [->](f)--(1);
     \draw [->](f)--(E);
   \draw [->](c)--(C);
    \draw [->](d)--(C);
    \draw [->](d)--(D);
    \draw [->](e)--(D);
    \draw [->](e)--(3);
    \draw [->](C)--(g);
     \draw [->](g)--(E);
     \draw [->](E)--(2);
     \draw [->](g)--(F);
       \draw [->](F)--(5);
      \draw [->](h)--(F);
      \draw [->](h)--(4);
    \draw [->](D)--(h);
  \end{tikzpicture}
\caption{\label{fig:Nn} Maximal sbTSTC phylogenetic networks with 3, 4, and 5 leaves.}
\end{figure}

Now, $N_3$ has 3 internal tree nodes and 2 hybrid nodes, and the
construction of $N_{n+1}$ from $N_n$ adds 3 internal tree nodes and 2
hybrid nodes. It is evident, then, that each $N_n$ has $3(n-2)$
internal tree nodes and $2(n-2)$ hybrid nodes. 
\end{exm}

\section{The mu-representation}
\label{sec:mu-representation}

In \cite{cardona.ea:07b} we introduced the $\mu$-representation for a
different class 
of phylogenetic networks, the so-called \emph{tree-child} phylogenetic
networks, those networks where every internal node has at least one child
that is a tree node.
We remark that the tree-child condition is more restrictive than the
tree-sibling one; nevertheless, the additional condition of time
consistency that we use here
makes that none of the two classes is contained in the other
one.

In this section we review the definition of
the $\mu$-representation of phylogenetic networks, and we will prove
later that this representation characterizes a sbTSTC phylogenetic
network, up to 
isomorphism. 

Let $N=(V,E)$ be a phylogenetic network on the set
$S=\{1,\dots,n\}$. For each node $u$ of $N$, we consider its
\emph{$\mu$-vector},
$$\mu(u)=(m_1(u),\dots,m_n(u)),$$
where $m_i(u)$ is the number of different paths from $u$ to the leaf
$i$. Moreover, we define the $\mu$-representation of $N$, $\mu(N)$, as
the multiset
$$\mu(N)=\{\mu(u)\mid u\in V\},$$
with each element appearing as many times as the number of
different nodes having it as its $\mu$-vector. 

For each leaf $i$, we have that
its $\mu$-vector is 
$\mu(i)=\delta(i)$, with $\delta(i)$ the
vector with $0$ at each position, except at its $i$-th position,
where it is $1$. As for the other nodes, we have that
$\mu(u)=\sum_{v_k}\mu(v_k)$, where the sum ranges over the set of
children of $u$ \cite[Lemma~4]{cardona.ea:07b}. This
property allows for the computation of $\mu(N)$ in polynomial time
(see Section~\ref{sec:comp-aspects} below).

\begin{exm}
  Consider the sbTSTC phylogenetic network in
Fig.~\ref{fig:exm-TSTC}. In Table~\ref{tab:mu-exm-TSTC} we give its
$\mu$-representation, except for the leaves, whose $\mu$-vector is
trivial.
\begin{table}
  \centering
  \caption{\label{tab:mu-exm-TSTC}$\mu$-representation of the network in Fig.~\ref{fig:exm-TSTC}.}
  \begin{tabular}{|c|c|}
    \hline
    node & $\mu$-vector \\ \hline
    $r$ & $(1,2,2,1)$ \\ \hline
    $u$ & $(1,1,0,0)$ \\ \hline
    $v$ & $(0,1,1,0)$ \\ \hline
    $w$ & $(0,0,1,1)$ \\ \hline
    $A$ & $(0,1,0,0)$ \\ \hline
    $B$ & $(0,0,1,0)$ \\ \hline
  \end{tabular}
\end{table}
\end{exm}

In the next section we will introduce a set of
decomposition/re\-con\-struc\-tion 
procedures for sbTSTC phylogenetic networks.
It will turn out that the
application conditions for these procedures
can be read from
the $\mu$-representation of the network.

\begin{lem}
  Let $N$ be a sbTSTC phylogenetic network, $i,j$ a pair of leaves,
  and let $u$ be the parent of $i$. Then $j$ is sibling or
  quasi-sibling of $i$ if, and only if:
  \begin{enumerate}
  \item $\mu(u)$ is minimal in the set
    $$M=\{\mu\in\mu(N)\mid \mu\ge\delta(i)+\delta(j)\}.$$
  \item The multiset
    $$M_i=\{\mu\in\mu(N)\mid\mu(u)>\mu\ge\delta(i)\}$$
    is equal to $\{\delta(i)\}$.
  \item The multiset
    $$M_j=\{\mu\in\mu(N)\mid\mu(u)>\mu\ge\delta(j)\}$$
    is equal to $\{\delta(j)\}$ 
    (when $j$ is sibling of $i$) or to $\{\delta(j),\delta(j)\}$
    (when $j$ is quasi-sibling of $i$).
  \end{enumerate}
\end{lem}

\begin{proof}
  Let us assume that $j$ is sibling or quasi-sibling of $i$.
  In either case, both $i$ and $j$ are
  descendants of $u$, so that  $\mu(u)\in M$. Now, for any other node
  $w$ with $\mu(w)\in M$, we have that $w\neq i$ and it is an ancestor
  of $i$,
  hence it is also an ancestor of $u$, and therefore $\mu(w)\ge
  \mu(u)$; hence, $\mu(u)$ is minimal in $M$.
  Moreover, the only $\mu$-vector in $M_i$ is $\delta(i)$,
  with multiplicity $1$,
  because the only ancestor of $i$ that is a non-trivial descendant of
  $u$ is the leaf $i$ itself.
  The situation for $M_j$ is analogous, taking into account that 
  $M_j$ contains a second copy of
  $\delta(j)$ in the case that the parent of $j$ is hybrid.

  As for the converse, let us assume that for a node $w$, its
  $\mu$-vector is minimal in $M$. Note that, since a hybrid node and
  its single child (a tree node) have the same $\mu$-vector, we can
  assume that $w$ is a tree node. Because of the definition of $M$,
  we have that $w$ is an ancestor of both $i$ and $j$. Now, if some
  child $v$ of $w$ were an ancestor of both $i$ and $j$, we would
  have that $\mu(w)>\mu(v)\ge\delta(i)+\delta(j)$, against our
  assumption on the minimality of $\mu(w)$ in $M$. Therefore, $w$ has
  two children $v_i,v_j$ such that $v_i$ is ancestor of $i$ (but not
  of $j$) and $v_j$ is ancestor of $j$ (but not of $i$). Then,
  $\mu(v_i)\in M_i$ and, by the uniqueness of the element in $M_i$, we
  have that $v_i=i$, and it follows that $w$ is the parent of $i$,
  that is,
  $w=u$.
  Symmetrically, we have that $v_j\in M_j$. Now,
  two situations may arise: first, if the multiplicity of $\delta(j)$ in
  $M_j$ is one, then $v_j=j$ and $j$ is a sibling of $i$; second, if
  this multiplicity is two, then $v_j$ must be a hybrid node whose
  single child is $j$, hence $j$ is quasi-sibling of $i$.
  
\end{proof}

\begin{lem}
  Let $N$ be a sbTSTC phylogenetic network.
  Let $j$ be a leaf sibling or quasi-sibling of another leaf
  $i$, and let $u$ be the parent of $i$. Then, $\outdeg(u)=2$ if, and only
  if, $\mu(u)=\delta(i)+\delta(j)$.
\end{lem}

\begin{proof}
  Note that with the assumptions made,
  and by the previous lemma, we have that
  $\mu(u)\ge\delta(i)+\delta(j)$. Now, the
  equality holds if, and only if, $u$ has no other children apart from
  $i$ and $j$ (in case that $j$ is sibling of $i$) or the hybrid
  parent of $j$ (in case that $j$ is quasi-sibling of $i$).
  
\end{proof}

For future reference, we gather these last results into the following
proposition. 

\begin{prop}\label{thm:sibligns-with-mu}
  Let $N$ be a sbTSTC phylogenetic network. The following properties
  can be decided from the knowledge of $\mu(N)$:
  \begin{enumerate}
  \item Two leaves are siblings, or not.
  \item A leaf is quasi-sibling of another one, or not.
  \item A leaf is sibling or quasi-sibling of another leaf, and the
    parent of the latter has out-degree $2$, or greater than $2$.
  \end{enumerate}
\end{prop}

\section{The reduction procedures}\label{sec:reduction-procedure}

We now introduce four reduction procedures that
decrease either the number of leaves or of hybrid nodes in a sbTSTC
phylogenetic network.

\subsubsection*{The $\Tred$ reduction.}

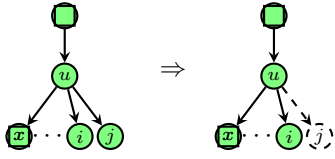
\begin{figure}
  \centering
  \begin{tikzpicture}[thick,>=stealth,scale=0.4,baseline=0pt]
    \draw (0,2) node[tre] (xx) {};
    \draw (0,2) node[hybn] (xx) {};
    \draw (0,0) node[tre] (u) {}; \etq{u}
    \draw (.5,-2) node[tre] (i) {}; \etq {i}
    \draw (1.5,-2) node[tre] (j) {}; \etq {j}
    \draw (-0.5,-2) node {$\cdots$};
    \draw (-1.5,-2) node[tre] (x) {}; \etq {x}
    \draw (-1.5,-2) node[hybn] (x) {}; \etq {x}
    \draw [->](xx)--(u);
    \draw [->](u)--(i);
    \draw [->](u)--(j);
    \draw [->](u)--(x);
  \end{tikzpicture}
  \quad$\Rightarrow$\quad
  \begin{tikzpicture}[thick,>=stealth,scale=0.4,baseline=0pt]
    \draw (0,2) node[tre] (xx) {};
    \draw (0,2) node[hybn] (xx) {};
    \draw (0,0) node[tre] (u) {}; \etq{u}
    \draw (.5,-2) node[tre] (i) {}; \etq {i}
    \draw (1.5,-2) node[tre,dashed,fill=white] (j) {}; \etq {j}
    \draw (-0.5,-2) node {$\cdots$};
    \draw (-1.5,-2) node[tre] (x) {}; \etq {x}
    \draw (-1.5,-2) node[hybn] (x) {}; \etq {x}
    \draw [->](xx)--(u);
    \draw [->](u)--(i);
    \draw [->,dashed,fill=white](u)--(j);
    \draw [->](u)--(x);
  \end{tikzpicture}
  \caption{\label{fig:T-reduction}The $\Tred$ reduction.}
\end{figure}

Let $N$ be a sbTSTC phylogenetic network on $S$,
$i,j$ two sibling leaves, $u$ their common parent, and assume
that $\outdeg(u)>2$. The DAG $N_{\Tred(i,j)}$ is
obtained by removing from $N$ the leaf $j$ and its incoming arc; see
Fig.~\ref{fig:T-reduction}.

It is easy to check that the obtained DAG is a sbTSTC
phylogenetic network on $S\setminus\{j\}$. Indeed, if the removed node
$j$ were a sibling of some hybrid node $x$, then $i$ would still be a
tree node sibling of $x$ in $N_{\Tred(i,j)}$, hence the tree-sibling condition is
preserved. Also, the time consistency and semi-binarity conditions
are trivially preserved.

Note that, given $N_{\Tred(i,j)}$, we can reconstruct $N$,
up to isomorphism,
by simply adding
the leaf $j$ and an arc from the parent of $i$ to $j$.

Note also that
the $\mu$-representation of $N_{\Tred(i,j)}$ can be easily obtained from
that of $N$. Indeed, for any node $u$ (except for the deleted leaf,
which implies removing $\delta(j)$ from $\mu(N)$) we
have that its $\mu$-vector in the reduced network is the same that in
the original network but with the $j$-th component removed.

\subsubsection*{The $\TRred$ reduction.}

\begin{figure}
\centering
  \begin{tikzpicture}[thick,>=stealth,scale=0.4,baseline=0pt]
    \draw (0,2) node[tre] (xx) {};
    \draw (0,2) node[hybn] (xx) {};
    \draw (0,0) node[tre] (u) {}; \etq{u}
    \draw (-.75,-2) node[tre] (i) {}; \etq {i}
    \draw (.75,-2) node[tre] (j) {}; \etq {j}
    \draw [->](xx)--(u);
    \draw [->](u)--(i);
    \draw [->](u)--(j);
  \end{tikzpicture}
  \quad$\Rightarrow$\quad
  \begin{tikzpicture}[thick,>=stealth,scale=0.4,baseline=0pt]
    \draw (0,2) node[tre] (xx) {};
    \draw (0,2) node[hybn] (xx) {};
    \draw (0,0) node[tre,dashed,fill=white] (u) {}; \etq{u}
    \draw (-.75,-2) node[tre] (i) {}; \etq {i}
    \draw (.75,-2) node[tre,dashed,fill=white] (j) {}; \etq {j}
    \draw [->,dashed](xx)--(u);
    \draw [->](xx) to [bend right] (i);
    \draw [->,dashed](u)--(i);
    \draw [->,dashed](u)--(j);
  \end{tikzpicture}
  \caption{\label{fig:TR-reduction}The $\TRred$ reduction.}
\end{figure}
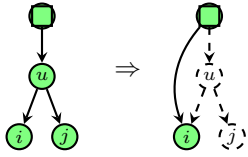

Let $N$ be a sbTSTC phylogenetic network on $S$,
$i,j$ two sibling leaves, $u$ their common parent, and assume
that $\outdeg(u)=2$.
Suppose also that $N$ is not a tree with two leaves, which
is equivalent to have that $u$ is not the root of $N$.
The DAG $N_{\TRred(i,j)}$ is
obtained by removing from $N$ the leaf $j$ and its incoming arc, and
collapsing the created elementary path into a single arc; see
Fig.~\ref{fig:TR-reduction}. 

As in the previous case, the resulting network is a
sbTSTC phylogenetic network on $S\setminus\{j\}$.
Indeed, if the node $u$ in $N$ is sibling of a hybrid node $w$, then
in the obtained network $N_{\TRred(i,j)}$ the leaf $i$ is a sibling of $w$.

Analogously to the previous case, given $N_{\TRred(i,j)}$, we can
reconstruct $N$ up to isomorphism 
by simply adding 
the leaf $j$,
splitting the arc with head $i$ by introducing an intermediate node
$u$, and adding 
an arc from $u$ to $j$.

Moreover, the $\mu$-representation of $N_{\TRred(i,j)}$ can be easily
obtained from 
that of $N$. The procedure is analogous to the previous case, taking
into account that we have also to remove from $\mu(N)$ a node with
$\mu$-vector equal to $\delta(i)+\delta(j)$.

\subsubsection*{The $\Hred$ reduction.}

\begin{figure}
\centering
  \begin{tikzpicture}[thick,>=stealth,scale=0.4,baseline=0pt,xscale=0.9]
    \draw (0,2) node[tre] (xx) {};
    \draw (0,2) node[hybn] (xx) {};
    \draw (0,0) node[tre] (u) {}; \etq{u}
    \draw (1.5,0) node[hyb] (v) {}; \etq v
    \draw (3,0) node[tre] (yy) {}; 
    \draw (0,-2) node[tre] (i) {}; \etq {i}
    \draw (1.5,-2) node[tre] (j) {}; \etq {j}
    \draw (-0.8,-2) node {$\scriptstyle\cdots$};
    \draw (-2,-2) node[tre] (x) {}; 
    \draw (-2,-2) node[hybn] (x) {}; 
    \draw [->](xx)--(u);
    \draw [->](u)--(i);
    \draw [->](u)--(v);
    \draw [->](u)--(x);
    \draw [->](yy)--(v);
    \draw [->](v)--(j);
  \end{tikzpicture}
  \quad$\Rightarrow$\quad
  \begin{tikzpicture}[thick,>=stealth,scale=0.4,baseline=0pt,xscale=0.9]
    \draw (0,2) node[tre] (xx) {};
    \draw (0,2) node[hybn] (xx) {};
    \draw (0,0) node[tre] (u) {}; \etq{u}
    \draw (1.5,0) node[hyb,dashed,fill=white] (v) {}; \etq v
    \draw (3,0) node[tre] (yy) {}; 
    \draw (0,-2) node[tre] (i) {}; \etq {i}
    \draw (1.5,-2) node[tre] (j) {}; \etq {j}
    \draw (-0.8,-2) node {$\scriptstyle\cdots$};
    \draw (-2,-2) node[tre] (x) {}; 
    \draw (-2,-2) node[hybn] (x) {}; 
    \draw [->](xx)--(u);
    \draw [->](u)--(i);
    \draw [->](u)--(x);
    \draw [->,dashed](u)--(v);
    \draw [->,dashed](yy)--(v);
    \draw [->](yy) to [bend right](j);
    \draw [->,dashed](v)--(j);
  \end{tikzpicture}
  \caption{\label{fig:H-reduction}The $\Hred$ reduction.}
\end{figure}
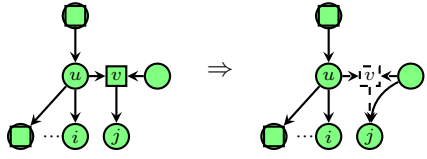

Let $N$ be a sbTSTC phylogenetic network on $S$,
$j$ a leaf quasi-sibling of another leaf $i$, $u$ the parent of $i$,
$v$ the parent of $j$,
and assume
that $\outdeg(u)>2$. The DAG $N_{\Hred(i,j)}$ is
obtained by removing from $N$ the
arc $(u,v)$ and 
collapsing the resulting elementary path with intermediate node $v$ into
a single arc; see Fig.~\ref{fig:H-reduction}.

Since we have only removed a hybrid node of $N$, when collapsing the
elementary path,
it is straightforward to check that the obtained DAG is a sbTSTC
phylogenetic network on $S$.

Now, given $N_{\Hred(i,j)}$, we can reconstruct $N$ up to isomorphism
by simply 
splitting the arc with head $j$ by introducing an intermediate node
$v$, and adding 
an arc from the parent of $i$ to $v$.

Note that the $\mu$-representation of $N_{\Hred(i,j)}$ can be easily
obtained from 
that of $N$. Namely,
for every node $x$ (except for the removed hybrid node, which implies
removing one copy of $\delta(j)$ from $\mu(N)$) we have that
if $\mu_N(x)=(m_1(x),\dots,m_n(x))$, then
$\mu_{N_{\Hred(i,j)}}(x)=(m'_1(x),\dots,m'_n(x))$ with
$$m'_k(x)=
\begin{cases}
  m_k(x)&\text{if $k\neq j$,}\\
  m_j(x)-m_i(x)&\text{if $k=j$.}
\end{cases}
$$
This follows from the fact that we have only removed the paths $x\pathgr j$
that pass through the parent of $i$, which are in bijection with the
paths $x\pathgr i$.

\subsubsection*{The $\HRred$ reduction.}

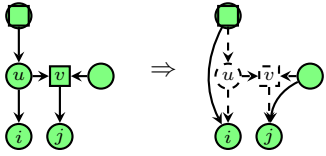
\begin{figure}
  \centering
  \begin{tikzpicture}[thick,>=stealth,scale=0.4,xscale=0.9,baseline=0pt]
    \draw (0,2) node[tre] (xx) {};
    \draw (0,2) node[hybn] (xx) {};
    \draw (0,0) node[tre] (u) {}; \etq{u}
    \draw (1.5,0) node[hyb] (v) {}; \etq v
    \draw (3,0) node[tre] (yy) {}; 
    \draw (0,-2) node[tre] (i) {}; \etq {i}
    \draw (1.5,-2) node[tre] (j) {}; \etq {j}
    \draw [->](xx)--(u);
    \draw [->](u)--(i);
    \draw [->](u)--(v);
    \draw [->](yy)--(v);
    \draw [->](v)--(j);
  \end{tikzpicture}
  \quad$\Rightarrow$\quad
  \begin{tikzpicture}[thick,>=stealth,scale=0.4,xscale=0.9,baseline=0pt]
    \draw (0,2) node[tre] (xx) {};
    \draw (0,2) node[hybn] (xx) {};
    \draw (0,0) node[tre,dashed,fill=white] (u) {}; \etq{u}
    \draw (1.5,0) node[hyb,dashed,fill=white] (v) {}; \etq v
    \draw (3,0) node[tre] (yy) {}; 
    \draw (0,-2) node[tre] (i) {}; \etq {i}
    \draw (1.5,-2) node[tre] (j) {}; \etq {j}
    \draw [->](xx) to [bend right] (i);
    \draw [->,dashed](xx)--(u);
    \draw [->,dashed](u)--(i);
    \draw [->,dashed](u)--(v);
    \draw [->,dashed](yy)--(v);
    \draw [->](yy) to [bend right](j);
    \draw [->,dashed](v)--(j);
  \end{tikzpicture}
  \caption{\label{fig:HR-reduction}The $\HRred$ reduction.}
\end{figure}

Let $N$ be a sbTSTC phylogenetic network on $S$,
$j$ a leaf quasi-sibling of another leaf $i$, $u$ the parent of $i$,
$v$ the parent of $j$,
and assume
that $\outdeg(u)=2$. The DAG $N_{\HRred(i,j)}$ is
obtained by removing from $N$ the
arc $(u,v)$ and 
collapsing the created elementary paths with respective intermediate
nodes $u$ and $v$ into
single arcs; see Fig.~\ref{fig:HR-reduction}.

The fact that the obtained DAG is a sbTSTC
phylogenetic network on $S$ follows as in the previous cases.

Also, given $N_{\HRred(i,j)}$, we can reconstruct $N$ by simply
splitting the arcs with respective heads $i,j$ by introducing
intermediate nodes 
$u,v$,
and adding 
an arc from $u$ to $v$.

Moreover, the $\mu$-representation of $N_{\HRred(i,j)}$ can be also
obtained from 
that of $N$. The procedure is the same as in the last case, taking
into account that we have also to remove from $\mu(N)$ a node with
$\mu$-vector equal to $\delta(i)+\delta(j)$.

\begin{exm}
  In Fig.~\ref{fig:reduction-TSTC} we show 
a sequence of reduction processes that, applied to the network in
Fig.~\ref{fig:exm-TSTC}, 
reduce it to a 
tree with two leaves.

\begin{figure}[t]
  \centering
  \begin{tikzpicture}[thick,>=stealth,scale=0.5,baseline=0pt]
    \draw(0,0) node[tre] (r) {}; \etq r;
    \draw(-2,-1) node[tre] (u) {}; \etq u;
    \draw(0,-1) node[tre] (v) {}; \etq v;
    \draw(2,-1) node[tre] (w) {}; \etq w;
    \draw(-1,-1) node[hyb] (A) {}; \etq A;
    \draw(1,-1) node[hyb] (B) {}; \etq B;
    \draw(-2,-2) node[tre] (1) {}; \etq 1;
    \draw(-1,-2) node[tre] (2) {}; \etq 2;
    \draw(1,-2) node[tre] (3) {}; \etq 3;
    \draw(2,-2) node[tre] (4) {}; \etq 4;
    \draw[->] (r)--(u);
    \draw[->] (r)--(v);
    \draw[->] (r)--(w);
    \draw[->] (u)--(A);
    \draw[->] (v)--(A);
    \draw[->] (v)--(B);
    \draw[->] (w)--(B);
    \draw[->] (u)--(1);
    \draw[->] (A)--(2);
    \draw[->] (B)--(3);
    \draw[->] (w)--(4);
  \end{tikzpicture}
  \begin{tikzpicture}[baseline=0pt,scale=0.5]
    \draw(0,0) {};
    \draw(0,-1) node {$\stackrel{\HRred(1,2)}{\Rightarrow}$};
    \draw(0,-2) {};
  \end{tikzpicture}
  \begin{tikzpicture}[thick,>=stealth,scale=0.5,baseline=0pt]
    \draw(0,0) node[tre] (r) {}; \etq r;
    \draw(0,-1) node[tre] (v) {}; \etq v;
    \draw(2,-1) node[tre] (w) {}; \etq w;
    \draw(1,-1) node[hyb] (B) {}; \etq B;
    \draw(-2,-2) node[tre] (1) {}; \etq 1;
    \draw(-1,-2) node[tre] (2) {}; \etq 2;
    \draw(1,-2) node[tre] (3) {}; \etq 3;
    \draw(2,-2) node[tre] (4) {}; \etq 4;
    \draw[->] (r)--(1);
    \draw[->] (r)--(v);
    \draw[->] (r)--(w);
    \draw[->] (v)--(2);
    \draw[->] (v)--(B);
    \draw[->] (w)--(B);
    \draw[->] (B)--(3);
    \draw[->] (w)--(4);
  \end{tikzpicture}
  \begin{tikzpicture}[baseline=0pt,scale=0.5]
    \draw(0,0) {};
    \draw(0,-1) node {$\stackrel{\HRred(4,3)}{\Rightarrow}$};
    \draw(0,-2) {};
  \end{tikzpicture}
  \begin{tikzpicture}[thick,>=stealth,scale=0.5,baseline=0pt]
    \draw(0,0) node[tre] (r) {}; \etq r;
    \draw(0,-1) node[tre] (v) {}; \etq v;
    \draw(-2,-2) node[tre] (1) {}; \etq 1;
    \draw(-1,-2) node[tre] (2) {}; \etq 2;
    \draw(1,-2) node[tre] (3) {}; \etq 3;
    \draw(2,-2) node[tre] (4) {}; \etq 4;
    \draw[->] (r)--(1);
    \draw[->] (r)--(v);
    \draw[->] (r)--(4);
    \draw[->] (v)--(2);
    \draw[->] (v)--(3);
  \end{tikzpicture}
  \begin{tikzpicture}[baseline=0pt,scale=0.5]
    \draw(0,0) {};
    \draw(0,-1) node {$\stackrel{\TRred(2,3)}{\Rightarrow}$};
    \draw(0,-2) {};
  \end{tikzpicture}
  \\
  \begin{tikzpicture}[thick,>=stealth,scale=0.5,baseline=0pt]
    \draw(0,0) node[tre] (r) {}; \etq r;
    \draw(-2,-2) node[tre] (1) {}; \etq 1;
    \draw(-1,-2) node[tre] (2) {}; \etq 2;
    \draw(1,-2) node[tre] (4) {}; \etq 4;
    \draw[->] (r)--(1);
    \draw[->] (r)--(2);
    \draw[->] (r)--(4);
  \end{tikzpicture}
  \begin{tikzpicture}[baseline=0pt,scale=0.5]
    \draw(0,0) {};
    \draw(0,-1) node {$\stackrel{\Tred(1,2)}{\Rightarrow}$};
    \draw(0,-2) {};
  \end{tikzpicture}
  \begin{tikzpicture}[thick,>=stealth,scale=0.5,baseline=0pt]
    \draw(0,0) node[tre] (r) {}; \etq r;
    \draw(-1,-2) node[tre] (1) {}; \etq 1;
    \draw(1,-2) node[tre] (4) {}; \etq 4;
    \draw[->] (r)--(1);
    \draw[->] (r)--(4);
  \end{tikzpicture}
  \caption{\label{fig:reduction-TSTC}Reduction processes for network in Fig.~\ref{fig:exm-TSTC}.}
\end{figure}
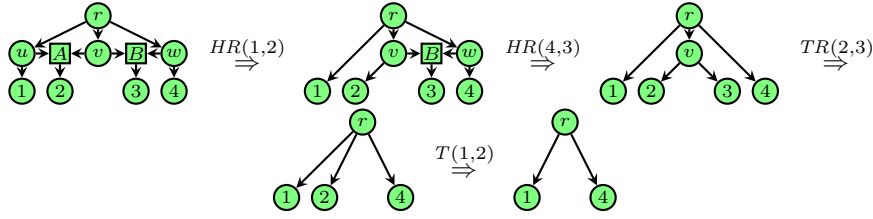
\end{exm}

\begin{rem}
  The construction given in Example~\ref{ex:maximum} for the networks
  with maximal number of nodes
  can also be described in terms of the reductions (or rather their
  inverses) we have defined.
  Indeed,
  $N_{n+1}$ can also be described as the network obtained from $N_n$
  by application of the
  inverses of the reductions $\TRred(2,n+1)$, $\HRred(1,2)$, and
  $\HRred(n,n+1)$ (in this order).
\end{rem}

\section{The mu-distance}

For any pair of phylogenetic networks $N_1,N_2$ on the same set
of leaves, let
$$d_\mu(N_1,N_2)=|\mu(N_1)\triangle \mu(N_2)|,$$
where both the symmetric difference and the cardinality operator refer
to multisets.

Our main result in this paper is that this mapping $d_\mu$ 
gives a distance on the class of sbTSTC phylogenetic networks
on a given set $S$ of taxa. We remark that $d_\mu$ is also a distance
on the set of tree-child phylogenetic networks on $S$
and, in particular, on phylogenetic trees,
where it coincides with the Robinson-Foulds distance
\cite{cardona.ea:07b}.

\begin{thm}
  Let $N_1,N_2,N_3$ be sbTSTC phylogenetic networks on the same set
  of taxa. Then:
  \begin{enumerate}
  \item $d_\mu(N_1,N_2)\ge 0$,
  \item $d_\mu(N_1,N_2)=0$ if, and only if, $N_1\cong N_2$,
  \item $d_\mu(N_1,N_2)=d_\mu(N_2,N_1)$,
  \item $d_\mu(N_1,N_3)\le d_\mu(N_1,N_2)+d_\mu(N_2,N_3)$.
  \end{enumerate}
\end{thm}

\begin{proof}
  Except for the second statement, the result follows from the
  properties of the symmetric difference of multisets.

  Also, if $N_1$ and $N_2$ are isomorphic, it follows from the
  definition of the $\mu$-representation that $\mu(N_1)$ and
  $\mu(N_2)$ are equal as multisets.
  
  We will prove
  the separation property ($d_\mu(N_1,N_2)=0$ implies that $N_1\cong N_2$)
  by induction on the number $n$ of leaves
  and the number $h$ of hybrid nodes.

  If $n\le 2$, which implies that $h=0$,
  the result is obvious, since there exists only
  two such sbTSTC phylogenetic networks, namely the rooted
  trees with 1 and 2 leaves. Also, when
  $h=0$,
  the networks are, in fact, trees and the separation property
  of the Robinson-Foulds distance implies that $N_1\cong N_2$.

  Let us assume that the result is proved for sbTSTC networks with at
  most $n-1\ge 2$ leaves, and with $n$ leaves and at most $h-1\ge0$ hybrid
  nodes. Let $N_1$, $N_2$ be sbTSTC phylogenetic networks with $n$
  leaves and $h$ hybrid nodes.
  Because of Lemma~\ref{thm:existance} there exists a pair of
  leaves $i,j$ 
  such that $j$ is a sibling of $i$ (respectively, $j$ is quasi-sibling of
  $i$) in $N_1$. Now since $\mu(N_1)=\mu(N_2)$, we can apply
  Proposition~\ref{thm:sibligns-with-mu} to get that $j$ is also a
  sibling (respectively, quasi-sibling) of $i$ in $N_2$. Moreover, also from
  Proposition~\ref{thm:sibligns-with-mu} it follows that the
  out-degree of the parent of $i$ in $N_1$ is equal to $2$ if, and
  only if, the out-degree of the parent of $i$ in $N_2$ is equal to
  $2$. From this, it follows that we can apply the same reduction
  to both networks;
  let $N'_1,N'_2$ the networks obtained from $N_1,N_2$ using this
  reduction. 
  Since the $\mu$-representation of the
  reductions depends only on the $\mu$-representation of the original
  network and the reduction procedure applied, we get that
  $\mu(N'_1)=\mu(N'_2)$.
  Since now $N'_1$ and $N'_2$ have less leaves
  or hybrid nodes than $N_1$ and $N_2$, it follows from the induction
  hypothesis that $N'_1\cong N'_2$. Finally, since we can recover
  up to isomorphisms
  the
  original networks from their reduced networks and the reductions applied, we
  conclude that $N_1\cong N_2$.
  
\end{proof}

The tight bounds found in Section~\ref{sec:tree-sibling-phn} for the
number of internal 
nodes in a sbTSTC phylogenetic network allow us to find the
\emph{diameter} of this class of phylogenetic networks with respect to
the $\mu$-distance, that is, the maximum of the distances between two
networks in this class. The interest of having a closed expression for
the diameter is that it allows to normalize the $\mu$-distance in
order to take values in the unit interval $[0,1]$ of real numbers.

\begin{prop}
\label{prop:diam}
The diameter of the class of sbTSTC phylogenetic networks with respect
to $d_{\mu}$  is  $0$ when $n\leq 2$, $9$ when $n=3$, and $10(n-2)$
when $n\geq 4$. 
\end{prop}

\begin{proof}
The assertion for $n\leq  2$ is straightforward: there is only one
sbTSTC phylogenetic network with one leaf and one sbTSTC phylogenetic
network with two leaves. As far as the assertion for $n=3$ goes, it can
be easily checked by means of the direct computation of all pairs of
distances: the largest distance is 9, and it is reached (up to
permutations of labels) only by the pair of networks depicted in
Fig.~\ref{fig:mu-max-3}. 
\begin{figure}[htb]
\centering 
  \begin{tikzpicture}[thick,>=stealth,scale=0.3]
    \draw (0,0) node[tre] (1) {}; \etq{1}
    \draw (2,0) node[tre] (2) {}; \etq{2}
     \draw (4,0) node[tre] (3) {}; \etq{3}
    \draw (2,2) node[hyb] (A) {};   
     \draw (4,2) node[hyb] (B) {}; 
    \draw (0,4) node[tre] (a) {};  
         \draw (4,4) node[tre] (b) {};  
             \draw (3,6) node[tre] (r) {};  
     \draw [->](r)--(a);
 \draw [->](r)--(b);
     \draw [->](a)--(1);
     \draw [->](a)--(A);
     \draw [->](a)--(B);
    \draw [->](b)--(A);
     \draw [->](b)--(B);
    \draw [->](A)--(2);
     \draw [->](B)--(3);
  \end{tikzpicture}
\quad
  \begin{tikzpicture}[thick,>=stealth,scale=0.3]
   \draw (2,0) node[tre] (1) {}; \etq{1}
    \draw (0,0) node[tre] (2) {}; \etq{2}
     \draw (4,0) node[tre] (3) {}; \etq{3}
    \draw (2,2) node[hyb] (A) {};   
     \draw (0,2.5) node[tre] (a) {}; 
    \draw (4,2.5) node[tre] (b) {};  
         \draw (2,4.5)   node[tre] (r) {};  
     \draw [->](r)--(a);
 \draw [->](r)--(b);
     \draw [->](a)--(2);
     \draw [->](a)--(A);
    \draw [->](b)--(A);
     \draw [->](b)--(3);
    \draw [->](A)--(1);
  \end{tikzpicture}
\caption{\label{fig:mu-max-3} A pair of sbTSTC phylogenetic networks with 3 leaves at maximum $\mu$-distance.}
\end{figure}
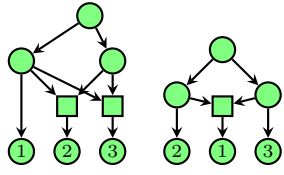

Finally, in the case $n\geq 4$, we know that a sbTSTC phylogenetic
network with $n$ leaves has at most $3(n-2)$ internal tree nodes and
$2(n-2)$ hybrid nodes, which gives an upper bound of $5(n-2)$ for the
total number of internal nodes. Now, the $\mu$-vector of the leaf $i$ is the
same in any sbTSTC phylogenetic network, and therefore the
$\mu$-distance between two sbTSTC phylogenetic networks is upper
bounded by the sum of their numbers of internal nodes. 

Combining these two upper bounds, we have that, for every pair of
sbTSTC phylogenetic networks with $n$ leaves $N$ and $N'$, 
$$
d_{\mu}(N,N')\leq 2\cdot 5(n-2)=10(n-2).
$$
It remains to display a pair of sbTSTC phylogenetic networks with $n$
leaves whose $\mu$-distance reaches this equality. Such a pair must
consist of two sbTSTC phylogenetic networks with $3(n-2)$ internal
tree nodes and $2(n-2)$ hybrid nodes each, and with disjoint sets of
$\mu$-vectors of internal nodes. 

One such pair is given by the network $N_n$ described in
Example~\ref{ex:maximum} and the network $N_n'$ obtained from $N_n$ by  
interchanging on the one hand the labels $1$ and $n$ and on the other
hand the labels $2$ and $3$. Fig.~\ref{fig:N5'} depicts $N_5'$ side by
side with $N_5$ to ease to spot the differences between these
networks. 

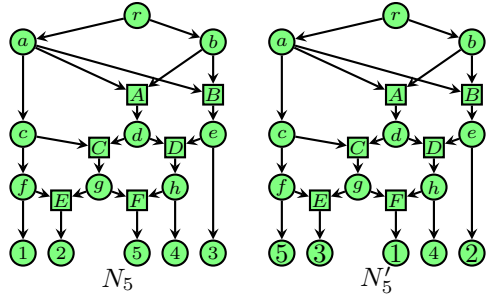
\begin{figure}[htb]
\centering 
  \begin{tikzpicture}[thick,>=stealth,xscale=0.25,yscale=0.35,baseline=0pt]
    \draw (0,-2) node[tre] (1) {}; \etq{1}
    \draw (2,-2) node[tre] (2) {}; \etq{2}
      \draw (6,-2) node[tre] (5) {}; \etq{5}
    \draw (8,-2) node[tre] (4) {}; \etq{4}
     \draw (10,-2) node[tre] (3) {}; \etq{3}
    \draw (2,0) node[hyb] (E) {};  \etq{E}
     \draw (6,0) node[hyb] (F) {}; \etq{F}
   \draw (0,0.5) node[tre] (f) {}; \etq{f}
         \draw (4,0.5) node[tre] (g) {};  \etq{g}
             \draw (8,0.5) node[tre] (h) {}; \etq{h}
    \draw (4,2) node[hyb] (C) {};  \etq{C}
     \draw (8,2) node[hyb] (D) {}; \etq{D}
    \draw (0,2.5) node[tre] (c) {}; \etq{c}
         \draw (6,2.5) node[tre] (d) {};  \etq{d}
             \draw (10,2.5) node[tre] (e) {}; \etq{e}
    \draw (6,4) node[hyb] (A) {};  \etq{A}
     \draw (10,4) node[hyb] (B) {}; \etq{B}
    \draw (0,6) node[tre] (a) {};  \etq{a}
     \draw (10,6) node[tre] (b) {}; \etq{b}
    \draw (6,7) node[tre] (r) {};  \etq{r}
       \draw (5,-3) node {$N_5$}; 
     \draw [->](r)--(a);
 \draw [->](r)--(b);
    \draw [->](a)--(A);
 \draw [->](a)--(c);
    \draw [->](a)--(B);
    \draw [->](b)--(A);
    \draw [->](b)--(B);
    \draw [->](A)--(d);
    \draw [->](B)--(e);
    \draw [->](c)--(f);
     \draw [->](f)--(1);
     \draw [->](f)--(E);
   \draw [->](c)--(C);
    \draw [->](d)--(C);
    \draw [->](d)--(D);
    \draw [->](e)--(D);
    \draw [->](e)--(3);
    \draw [->](C)--(g);
     \draw [->](g)--(E);
     \draw [->](E)--(2);
     \draw [->](g)--(F);
       \draw [->](F)--(5);
      \draw [->](h)--(F);
      \draw [->](h)--(4);
    \draw [->](D)--(h);
  \end{tikzpicture}
\quad  \begin{tikzpicture}[thick,>=stealth,xscale=0.25,yscale=0.35,baseline=0pt]
    \draw (0,-2) node[tre] (1) {}; \draw (1) node {$\textstyle 5$};
    \draw (2,-2) node[tre] (2) {}; \draw (2) node {$\textstyle 3$};
      \draw (6,-2) node[tre] (5) {}; \draw (5) node {$\textstyle 1$};
    \draw (8,-2) node[tre] (4) {}; \etq{4}
     \draw (10,-2) node[tre] (3) {}; \draw (3) node {$\textstyle 2$};
    \draw (2,0) node[hyb] (E) {};  \etq{E}
     \draw (6,0) node[hyb] (F) {}; \etq{F}
   \draw (0,0.5) node[tre] (f) {}; \etq{f}
         \draw (4,0.5) node[tre] (g) {};  \etq{g}
             \draw (8,0.5) node[tre] (h) {}; \etq{h}
    \draw (4,2) node[hyb] (C) {};  \etq{C}
     \draw (8,2) node[hyb] (D) {}; \etq{D}
    \draw (0,2.5) node[tre] (c) {}; \etq{c}
         \draw (6,2.5) node[tre] (d) {};  \etq{d}
             \draw (10,2.5) node[tre] (e) {}; \etq{e}
    \draw (6,4) node[hyb] (A) {};  \etq{A}
     \draw (10,4) node[hyb] (B) {}; \etq{B}
    \draw (0,6) node[tre] (a) {};  \etq{a}
     \draw (10,6) node[tre] (b) {}; \etq{b}
    \draw (6,7) node[tre] (r) {};  \etq{r}
       \draw (5,-3) node {$N_5'$}; 
     \draw [->](r)--(a);
 \draw [->](r)--(b);
    \draw [->](a)--(A);
 \draw [->](a)--(c);
    \draw [->](a)--(B);
    \draw [->](b)--(A);
    \draw [->](b)--(B);
    \draw [->](A)--(d);
    \draw [->](B)--(e);
    \draw [->](c)--(f);
     \draw [->](f)--(1);
     \draw [->](f)--(E);
   \draw [->](c)--(C);
    \draw [->](d)--(C);
    \draw [->](d)--(D);
    \draw [->](e)--(D);
    \draw [->](e)--(3);
    \draw [->](C)--(g);
     \draw [->](g)--(E);
     \draw [->](E)--(2);
     \draw [->](g)--(F);
       \draw [->](F)--(5);
      \draw [->](h)--(F);
      \draw [->](h)--(4);
    \draw [->](D)--(h);
  \end{tikzpicture}
\caption{\label{fig:N5'} Two sbTSTC phylogenetic networks with   5 leaves at maximum $\mu$-distance.}
\end{figure}

To prove that $N_n$ and $N_n'$ have disjoint sets of $\mu$-vectors of
internal nodes, let us start by studying the \emph{clusters} (that is,
the sets of descendant leaves) of their internal nodes. We shall
denote the cluster of a node $v$ in a network $N$ by $C_N(v)$, and we
shall say that such a cluster is \emph{internal} when $v$ is
internal. Note that if two nodes have different clusters, then they
must have different $\mu$-vectors. 

The construction of $N_n$ from $N_{n-1}$ changes its set of internal
clusters in the following way. On the one hand, every internal node of
$N_{n-1}$ survives in $N_n$  and its cluster is modified in the
following way: 
\begin{itemize}
\item $C_{N_{n-1}}(v)\subseteq C_{N_n}(v)$.

\item If  $1\in C_{N_{n-1}}(v)$, then 2 is added to $C_{N_n}(v)$.

\item If  $2\in C_{N_{n-1}}(v)$, then $n$  is added to $C_{N_n}(v)$.

\item If  $n-1\in C_{N_{n-1}}(v)$, then $n$  is added to $C_{N_n}(v)$.

\item No other leaf is added to  any cluster of an internal node.
\end{itemize}
On the other hand, this construction adds five new internal nodes with clusters
$$
\{1,2\},\{2\},\{2,n\},\{n\},\{n-1,n\}.
$$
Starting with the family of internal clusters  of $N_3$ and using
these rules, it is easy to prove by induction that the family of
internal clusters of $N_n$ is (up to repetitions) 
$$
\begin{array}{l}
\{1,2,3,4,\ldots,n\},\{2,3,4,\ldots,n\},\{3,4,\ldots,n\},
\{4,\ldots,n\}, \ldots,\{n-1,n\},\{n\},\\
\{1,2,5,6,\ldots,n\},\{1,2,6,\ldots,n\},\ldots,\{1,2,n-1,n\},
\{1,2,n\},\{1,2\},\\
\{2,5,6,\ldots,n\},\{2,6,\ldots,n\},\ldots,\{2,n-1,n\},\{2,n\},\{2\},\\
\{2,4,5,6,\ldots,n\}.
\end{array}
$$
Now, $N_n'$ is obtained from $N_n$ by interchanging $1$ with $n$ and
$2$ with $3$, and therefore 
the clusters of its internal nodes can be obtained from the clusters
of $N_n$ by applying this permutation. We conclude that the family of
internal clusters of $N'_n$ is (again, up to repetitions) 
$$
\begin{array}{l}
\{1,2,3,4,\ldots,n\},\{1,2,3,4,\ldots,n-1\},\{1,2,4,\ldots,n-1\},
\{1,4,\ldots,n-1\},\ldots,\{1,n-1\},\{1\},\\
\{1,3,5,6,\ldots,n\},\{1,3,6,\ldots,n\},\ldots,\{1,3,n-1,n\},
\{1,3,n\},\{3,n\},\\
\{1,3,5,6,\ldots,n-1\},\{1,3,6,\ldots,n-1\},\ldots,\{1,3,n-1\},
\{1,3\},\{3\},\\
\{1,3,4,5,6,\ldots,n-1\}.
\end{array}
$$
A simple inspection shows that only one cluster appears in both lists:
the whole $\{1,\ldots,n\}$. 
(Indeed, all internal clusters of $N_n$ contain the leaf $n$, except
$\{1,2\}$ and $\{2\}$. Now, on the one hand, the latter are not
internal clusters of $N_n'$ and, on the other hand, every internal
cluster in $N_n'$ containing $n$ also contains $1,3$, while no
internal cluster of $N_n$ other than  
$\{1,2,3,\ldots,n\}$ contains $1,3$.)

So, if a pair of internal nodes of $N_n$ and $N'_n$ have the same
$\mu$-vector, their clusters must be equal to $\{1,\ldots,n\}$.
Now, both $N_n$ and $N_n'$ have exactly two nodes with cluster
$\{1,\ldots,n\}$: the root and its out-degree 3 child $a$. The
$\mu$-vectors of $a$ or $r$ in $N_n$ are different from the
$\mu$-vectors of $a$ or $r$ in $N_n'$: in $N_n$, there is only one
path from $r$ and from $a$ to 1, while in $N_n'$ it is clear that
there is more than one such path (the parent of $1$ in
$N'$ is a hybrid node, and its two parents are descendants of
both $a$
and $r$).

Therefore, $N_n$ and $N'_n$ have disjoint sets of $\mu$-vectors of
internal nodes and their $\mu$-distance is $10(n-2)$.

\end{proof}

As discussed before, we can now define the \emph{normalized
  $\mu$-distance} as
$$\bar d_\mu(N_1,N_2)=\frac1{10(n-2)}d_\mu(N_1,N_2)$$
if the involved networks have $n>3$ leaves, or $\bar
d_\mu(N_1,N_2)=\frac19 d_\mu(N_1,N_2)$ if $n=3$. This way, $\bar d_\mu$
takes values in the interval $[0,1]$, and there exists pairs of
networks at maximum normalized distance $1$ for every number of
leaves. 

\begin{exm}
Consider now the phylogenetic networks in
Fig.~\ref{fig-exm-nakhleh}. The two networks $N_1,N_2$ are 
adapted from networks (a) and (b)
in \cite[Fig.~10]{nakhleh.ea:molbio07} (where we have substituted
the actual names of the species by integers identifying them); we
remark that the third 
one in the aforementioned paper and figure is isomorphic to the first
one. The phylogenetic tree $T$ depicted above is the underlying tree
from which 
both networks are obtained by adding edges corresponding to horizontal
gene transfer events. Both networks are binary and time consistent;
however, the first one is tree-child (hence tree-sibling) while the
second one is not tree-child, but it is tree-sibling. Also, the tree
can be considered a binary tree-sibling time consistent phylogenetic
network. Hence, we can compute their $\mu$-distances, obtaining that
the two networks are more similar to the underlying phylogenetic tree
that to each other:
$$
\begin{aligned}
  d_\mu(T,N_1)&=22, & \bar d_\mu(T,N_1)&\approx 0.169,\\
  d_\mu(T,N_2)&=32, & \bar d_\mu(T,N_2)&\approx 0.246,\\
  d_\mu(N_1,N_2)&=38, & \bar d_\mu(N_1,N_2)&\approx 0.292.
\end{aligned}
$$


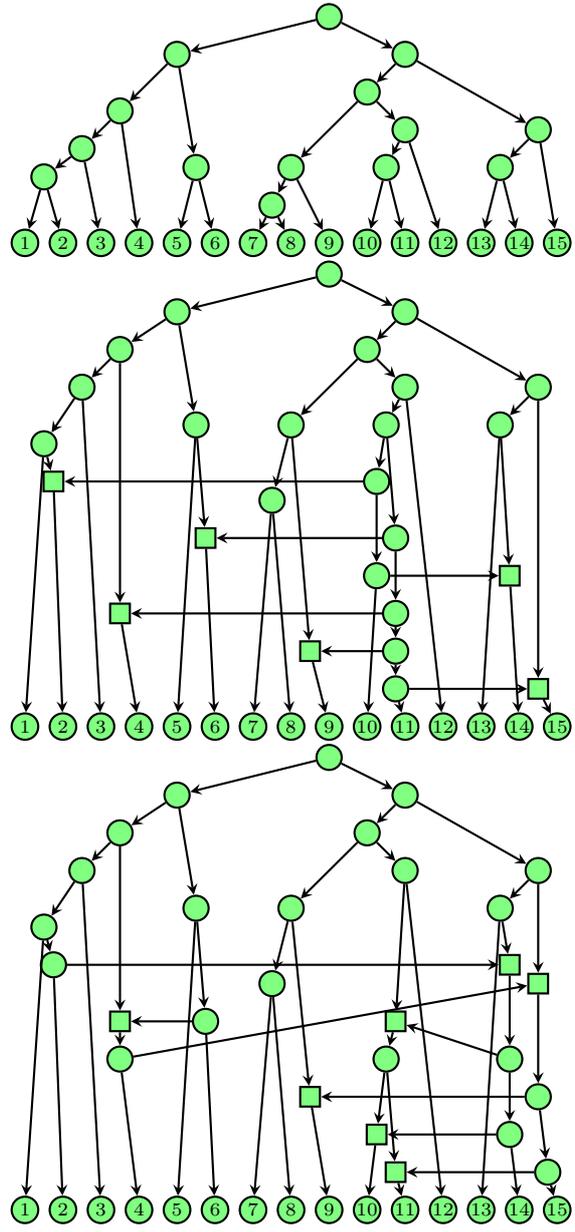
\begin{figure}
  \centering
  \begin{tikzpicture}[thick,>=stealth,xscale=0.5,yscale=0.5]
    \draw(9,0) node[tres] (r) {};
    \draw(11,-1) node[tres] (a) {};
    \draw(10,-2) node[tres] (b) {};
    \draw(5,-1) node[tres] (c) {};
    \draw(14.5,-3) node[tres] (d) {};
    \draw(11,-3) node[tres] (e) {};
    \draw(3.5,-2.5) node[tres] (f) {};
    \draw(13.5,-4) node[tres] (g) {};
    \draw(10.5,-4) node[tres] (h) {};
    \draw(8,-4) node[tres] (i) {};
    \draw(5.5,-4) node[tres] (j) {};
    \draw(2.5,-3.5) node[tres] (k) {};
    \draw(7.5,-5) node[tres] (l) {};
    \draw(1.5,-4.25) node[tres] (m) {};
    \draw(1,-6) node[trem] (1) {};
    \draw(2,-6) node[trem] (2) {};
    \draw(3,-6) node[trem] (3) {};
    \draw(4,-6) node[trem] (4) {};
    \draw(5,-6) node[trem] (5) {};
    \draw(6,-6) node[trem] (6) {};
    \draw(7,-6) node[trem] (7) {};
    \draw(8,-6) node[trem] (8) {};
    \draw(9,-6) node[trem] (9) {};
    \draw(10,-6) node[trem] (10) {};
    \draw(11,-6) node[trem] (11) {};
    \draw(12,-6) node[trem] (12) {};
    \draw(13,-6) node[trem] (13) {};
    \draw(14,-6) node[trem] (14) {};
    \draw(15,-6) node[trem] (15) {};
    \etq{1}\etq{2}\etq{3}\etq{4}\etq{5}\etq{6}\etq{7}\etq{8}\etq{9}\etq{10}\etq{11}\etq{12}\etq{13}\etq{14}\etq{15}
    \draw[->] (r)--(a);
    \draw[->] (r)--(c);
    \draw[->] (a)--(b);
    \draw[->] (a)--(d);
    \draw[->] (b)--(e);
    \draw[->] (b)--(i);
    \draw[->] (c)--(j);
    \draw[->] (c)--(f);
    \draw[->] (d)--(g);
    \draw[->] (d)--(15);
    \draw[->] (e)--(h);
    \draw[->] (e)--(12);
    \draw[->] (f)--(4);
    \draw[->] (f)--(k);
    \draw[->] (g)--(13);
    \draw[->] (g)--(14);
    \draw[->] (h)--(10);
    \draw[->] (h)--(11);
    \draw[->] (i)--(l);
    \draw[->] (i)--(9);
    \draw[->] (j)--(5);
    \draw[->] (j)--(6);
    \draw[->] (k)--(m);
    \draw[->] (k)--(3);
    \draw[->] (l)--(7);
    \draw[->] (l)--(8);
    \draw[->] (m)--(1);
    \draw[->] (m)--(2);
  \end{tikzpicture}
  \begin{tikzpicture}[thick,>=stealth,xscale=0.5,yscale=1]
    \draw(9,-2) node[tres] (r) {};
    \draw(11,-2.5) node[tres] (a) {};
    \draw(10,-3) node[tres] (b) {};
    \draw(5,-2.5) node[tres] (c) {};
    \draw(14.5,-3.5) node[tres] (d) {};
    \draw(11,-3.5) node[tres] (e) {};
    \draw(3.5,-3) node[tres] (f) {};
    \draw(13.5,-4) node[tres] (g) {};
    \draw(10.5,-4) node[tres] (h) {};
    \draw(8,-4) node[tres] (i) {};
    \draw(5.5,-4) node[tres] (j) {};
    \draw(2.5,-3.5) node[tres] (k) {};
    \draw(7.5,-5) node[tres] (l) {};
    \draw(1.5,-4.25) node[tres] (m) {};
    \draw(1,-8) node[trem] (1) {};
    \draw(2,-8) node[trem] (2) {};
    \draw(3,-8) node[trem] (3) {};
    \draw(4,-8) node[trem] (4) {};
    \draw(5,-8) node[trem] (5) {};
    \draw(6,-8) node[trem] (6) {};
    \draw(7,-8) node[trem] (7) {};
    \draw(8,-8) node[trem] (8) {};
    \draw(9,-8) node[trem] (9) {};
    \draw(10,-8) node[trem] (10) {};
    \draw(11,-8) node[trem] (11) {};
    \draw(12,-8) node[trem] (12) {};
    \draw(13,-8) node[trem] (13) {};
    \draw(14,-8) node[trem] (14) {};
    \draw(15,-8) node[trem] (15) {};
    \etq{1}\etq{2}\etq{3}\etq{4}\etq{5}\etq{6}\etq{7}\etq{8}\etq{9}\etq{10}\etq{11}\etq{12}\etq{13}\etq{14}\etq{15}
    \draw(10.75,-7.5) node[tres] (t1) {};
    \draw(14.5,-7.5) node[hyb] (h1) {};
    \draw(10.75,-7) node[tres] (t2) {};
    \draw(8.5,-7) node[hyb] (h2) {};
    \draw(10.75,-6.5) node[tres] (t3) {};
    \draw(3.5,-6.5) node[hyb] (h3) {};
    \draw(10.25,-6) node[tres] (t4) {};
    \draw(13.75,-6) node[hyb] (h4) {};
    \draw(10.25,-4.75) node[tres] (t5) {};
    \draw(1.75,-4.75) node[hyb] (h5) {};
    \draw(10.75,-5.5) node[tres] (t6) {};
    \draw(5.75,-5.5) node[hyb] (h6) {};
    \draw[->](r)--(a);
    \draw[->](r)--(c);
    \draw[->](a)--(b);
    \draw[->](a)--(d);
    \draw[->](b)--(e);
    \draw[->](b)--(i);
    \draw[->](c)--(j);
    \draw[->](c)--(f);
    \draw[->](d)--(g);
    \draw[->](d)--(h1);
    \draw[->](h1)--(15);
    \draw[->](e)--(h);
    \draw[->](e)--(12);
    \draw[->](f)--(h3);
    \draw[->](h3)--(4);
    \draw[->](f)--(k);
    \draw[->](g)--(13);
    \draw[->](g)--(h4);
    \draw[->](h4)--(14);
    \draw[->](h)--(t5);
    \draw[->](t5)--(t4);
    \draw[->](t4)--(10);
    \draw[->](h)--(t6);
    \draw[->](t6)--(t3);
    \draw[->](t3)--(t2);
    \draw[->](t2)--(t1);
    \draw[->](t1)--(11);
    \draw[->](i)--(l);
    \draw[->](i)--(h2);
    \draw[->](h2)--(9);
    \draw[->](j)--(5);
    \draw[->](j)--(h6);
    \draw[->](h6)--(6);
    \draw[->](k)--(m);
    \draw[->](k)--(3);
    \draw[->](l)--(7);
    \draw[->](l)--(8);
    \draw[->](m)--(1);
    \draw[->](m)--(h5);
    \draw[->](h5)--(2);
    \draw[->](t1)--(h1);
    \draw[->](t2)--(h2);
    \draw[->](t3)--(h3);
    \draw[->](t4)--(h4);
    \draw[->](t5)--(h5);
    \draw[->](t6)--(h6);
  \end{tikzpicture}
    \begin{tikzpicture}[thick,>=stealth,xscale=0.5]
    \draw(9,-2) node[tres] (r) {};
    \draw(11,-2.5) node[tres] (a) {};
    \draw(10,-3) node[tres] (b) {};
    \draw(5,-2.5) node[tres] (c) {};
    \draw(14.5,-3.5) node[tres] (d) {};
    \draw(11,-3.5) node[tres] (e) {};
    \draw(3.5,-3) node[tres] (f) {};
    \draw(13.5,-4) node[tres] (g) {};
    \draw(10.5,-6) node[tres] (h) {};
    \draw(8,-4) node[tres] (i) {};
    \draw(5.5,-4) node[tres] (j) {};
    \draw(2.5,-3.5) node[tres] (k) {};
    \draw(7.5,-5) node[tres] (l) {};
    \draw(1.5,-4.25) node[tres] (m) {};
    \draw(1,-8) node[trem] (1) {};
    \draw(2,-8) node[trem] (2) {};
    \draw(3,-8) node[trem] (3) {};
    \draw(4,-8) node[trem] (4) {};
    \draw(5,-8) node[trem] (5) {};
    \draw(6,-8) node[trem] (6) {};
    \draw(7,-8) node[trem] (7) {};
    \draw(8,-8) node[trem] (8) {};
    \draw(9,-8) node[trem] (9) {};
    \draw(10,-8) node[trem] (10) {};
    \draw(11,-8) node[trem] (11) {};
    \draw(12,-8) node[trem] (12) {};
    \draw(13,-8) node[trem] (13) {};
    \draw(14,-8) node[trem] (14) {};
    \draw(15,-8) node[trem] (15) {};
    \etq{1}\etq{2}\etq{3}\etq{4}\etq{5}\etq{6}\etq{7}\etq{8}\etq{9}\etq{10}\etq{11}\etq{12}\etq{13}\etq{14}\etq{15}
    \draw(14.75,-7.5) node[tres] (t1) {};
    \draw(10.75,-7.5) node[hyb] (h1) {};
    \draw(14.5,-6.5) node[tres] (t2) {};
    \draw(8.5,-6.5) node[hyb] (h2) {};
    \draw(13.75,-6) node[tres] (t3) {};
    \draw(10.75,-5.5) node[hyb] (h3) {};
    \draw(13.75,-7) node[tres] (t4) {};
    \draw(10.25,-7) node[hyb] (h4) {};
    \draw(1.75,-4.75) node[tres] (t5) {};
    \draw(13.75,-4.75) node[hyb] (h5) {};
    \draw(5.75,-5.5) node[tres] (t6) {};
    \draw(3.5,-5.5) node[hyb] (h6) {};
    \draw(3.5,-6) node[tres] (t7) {};
    \draw(14.5,-5) node[hyb] (h7) {};
    \draw[->](r)--(a);
\draw[->](r)--(c);
\draw[->](a)--(b);
\draw[->](a)--(d);
\draw[->](b)--(e);
\draw[->](b)--(i);
\draw[->](c)--(j);
\draw[->](c)--(f);
\draw[->](d)--(g);
\draw[->](d)--(h7);
\draw[->](h7)--(t2);
\draw[->](t2)--(t1);
\draw[->](t1)--(15);
\draw[->](e)--(h3);
\draw[->](h3)--(h);
\draw[->](e)--(12);
\draw[->](f)--(h6);
\draw[->](h6)--(t7);
\draw[->](t7)--(4);
\draw[->](f)--(k);
\draw[->](g)--(13);
\draw[->](g)--(h5);
\draw[->](h5)--(t3);
\draw[->](t3)--(t4);
\draw[->](t4)--(14);
\draw[->](h)--(h4);
\draw[->](h4)--(10);
\draw[->](h)--(h1);
\draw[->](h1)--(11);
\draw[->](i)--(l);
\draw[->](i)--(h2);
\draw[->](h2)--(9);
\draw[->](j)--(5);
\draw[->](j)--(t6);
\draw[->](t6)--(6);
\draw[->](k)--(m);
\draw[->](k)--(3);
\draw[->](l)--(7);
\draw[->](l)--(8);
\draw[->](m)--(1);
\draw[->](m)--(t5);
\draw[->](t5)--(2);
\draw[->](t1)--(h1);
\draw[->](t2)--(h2);
\draw[->](t3)--(h3);
\draw[->](t4)--(h4);
\draw[->](t5)--(h5);
\draw[->](t6)--(h6);
\draw[->](t7)--(h7);
  \end{tikzpicture}
  \caption{Tree $T$ (above) and networks $N_1$ (middle), $N_2$ (below)
    from \cite[Fig.~10]{nakhleh.ea:molbio07}. 
    \label{fig-exm-nakhleh}}
\end{figure}

\end{exm}

\section{Computational aspects}\label{sec:comp-aspects}

We have already mentioned in Section~\ref{sec:mu-representation} that
the $\mu$-representation of a phylogenetic network can be efficiently
computed by means of a simple bottom-up technique. Indeed, if we
define the \emph{height} of a node as the length of the longest path
starting in this node, we get a stratification of nodes. The nodes
with height $0$ are the leaves, and their $\mu$-vectors are trivially
computed. Assuming that we have computed the $\mu$-vectors of nodes up
to a given height $h$, we can compute the $\mu$-vector of a node at
height $h+1$ by simply adding up the $\mu$-vectors of its children, that
are already computed. If the network has $n$ leaves, $m$ nodes, and
the out-degree of tree nodes is bounded by $k<m$, the
cost of this computation is $O(kmn)=O(m^2n)$. In order to improve the
efficency of the computations of distances below, the
$\mu$-representation of the network is stored with the $\mu$-vectors
sorted in any total order, for
instance the 
lexicographic order; note that the computational cost of sorting the
$\mu$-representation is
$O(nm\log m)$; hence, the total cost of the computation and sorting is
still $O(m^2n)$.

Also, given two networks and their $\mu$-representations, their
$\mu$-distance can be computed efficiently. We can assume that the
$\mu$-vectors of each network are sorted as explained above.
Then, a simultaneous traversal of the
$\mu$-representation of both networks allows the computation of their
$\mu$-distance in $O(n(m_1+m_2))$, where $m_1,m_2$ are the number of
nodes of each of the networks.

We have implemented the computation of the $\mu$-representation of
networks and the $\mu$-distance between them in a Perl package
\cite{cardona.ea:08c}, part of the BioPerl bundle
\cite{stajich.ea:2002}. 

Note also that the reduction procedures introduced in
Section~\ref{sec:reduction-procedure} allow for the construction of
all semi-binary tree-sibling time consistent phylogenetic networks on
a given set of taxa. Indeed, as we have already proved, each such a
network can be reduced to a tree with two leaves by recursively
applying the reduction procedures. Since all these procedures are
reversible, we can effectively construct all networks. However, the
computational cost of this construction is high, since for obtaining
all the sbTSTC networks over a set $S$ of leaves with
$h$ hybrid nodes we need to recursively construct, first, all the networks
with set of leaves $S'\subset S$, $|S'|=|S|-1$ and $h$ hybrid nodes,
and, second, all those with set of leaves $S$ and $h-1$ hybrid nodes. 


The aforementioned Perl package contains a module to construct all
\emph{tree-child} phylogenetic networks on a given set of leaves. We
are working on a module that generates all sbTSTC phylogenetic
networks, which will be incorporated in the next release of the package.

\section{Counterexamples}
\label{sec:counterexamples}

When we have defined the class of sbTSTC phylogenetic networks, we
have remarked that the conditions imposed are necessary in order to
single out networks by means of its $\mu$-representation. In this
section we give examples of pairs of more general, non-isomorphic
networks but with 
the same $\mu$-representation.

\begin{figure}
  \centering
  \begin{tikzpicture}[thick,>=stealth,baseline=0pt,scale=0.5]
    \draw(0,0) node[tre] (a) {};
    \draw(-1,-1) node[tre] (b) {};
    \draw(0.5,-1) node[hyb] (c) {};
    \draw(1.5,-1) node[tre] (d) {};
    \draw(-1,-2) node[tre] (e) {};
    \draw(-1.5,-3) node[tre] (f) {};
    \draw(-0.5,-3) node[tre] (g) {};
    \draw(0.5,-3) node[tre] (h) {};
    \draw(1.5,-3) node[tre] (i) {};
    \draw(-1.5,-4.5) node[hyb] (j) {};
    \draw(-0.5,-4.5) node[hyb] (k) {};
    \draw(0.5,-4.5) node[hyb] (l) {};
    \draw(1.5,-4.5) node[hyb] (m) {};
    \draw(-1.5,-5.5) node[tre] (1) {};\etq1
    \draw(-0.5,-5.5) node[tre] (2) {};\etq2
    \draw(0.5,-5.5) node[tre] (3) {};\etq3
    \draw(1.5,-5.5) node[tre] (4) {};\etq4
    \draw[->](a)--(b);
    \draw[->](a)--(d);
    \draw[->](b)--(c);
    \draw[->](d)--(c);
    \draw[->](b)--(e);
    \draw[->](d)--(i);
    \draw[->](c)--(h);
    \draw[->](e)--(f);
    \draw[->](e)--(g);
    \draw[->](f)--(j);
    \draw[->](f)--(m);
    \draw[->](g)--(k);
    \draw[->](g)--(l);
    \draw[->](h)--(k);
    \draw[->](h)--(l);
    \draw[->](i)--(j);
    \draw[->](i)--(m);
    \draw[->](j)--(1);
    \draw[->](k)--(2);
    \draw[->](l)--(3);
    \draw[->](m)--(4);
  \end{tikzpicture}\qquad
  \begin{tikzpicture}[thick,>=stealth,baseline=0pt,scale=0.5]
    \draw(0,0) node[tre] (a) {};

    \draw(-0.5,-1) node[tre] (b) {};
    \draw(-1.5,-2) node[tre] (c) {};
    \draw(.5,-2) node[hyb] (d) {};
    \draw(1.5,-2) node[tre] (e) {};

    \draw(-1.5,-3) node[tre] (f) {};
    \draw(-0.5,-3) node[tre] (g) {};
    \draw(0.5,-3) node[tre] (h) {};
    \draw(1.5,-3) node[tre] (i) {};
    \draw(-1.5,-4.5) node[hyb] (j) {};
    \draw(-0.5,-4.5) node[hyb] (k) {};
    \draw(0.5,-4.5) node[hyb] (l) {};
    \draw(1.5,-4.5) node[hyb] (m) {};
    \draw(-1.5,-5.5) node[tre] (1) {};\etq1
    \draw(-0.5,-5.5) node[tre] (2) {};\etq2
    \draw(0.5,-5.5) node[tre] (3) {};\etq3
    \draw(1.5,-5.5) node[tre] (4) {};\etq4
    \draw[->](a)--(b);
    \draw[->](a)--(e);
    \draw[->](b)--(c);
    \draw[->](b)--(g);
    \draw[->](c)--(f);
    \draw[->](c)--(d);
    \draw[->](d)--(h);
    \draw[->](e)--(d);
    \draw[->](e)--(i);
    \draw[->](f)--(j);
    \draw[->](f)--(m);
    \draw[->](g)--(k);
    \draw[->](g)--(l);
    \draw[->](h)--(k);
    \draw[->](h)--(l);
    \draw[->](i)--(j);
    \draw[->](i)--(m);
    \draw[->](j)--(1);
    \draw[->](k)--(2);
    \draw[->](l)--(3);
    \draw[->](m)--(4);
  \end{tikzpicture}
  \caption{\label{fig:non-tree-sibling}Non tree-sibling, semi-binary time consistent networks with
    the same $\mu$-representation.}
\end{figure}
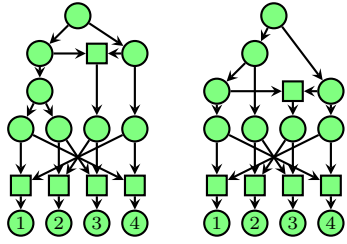

In Fig.~\ref{fig:non-tree-sibling} we give an example of a pair of
semi-binary time consistent networks not satisfying the tree-sibling
property, and having the same $\mu$-representation.

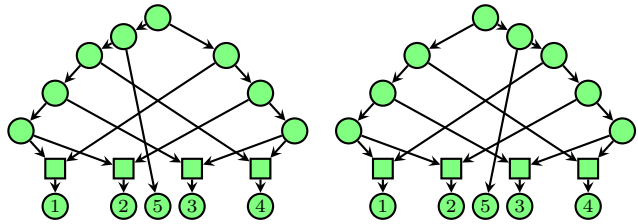
\begin{figure}
  \centering
  \begin{tikzpicture}[thick,>=stealth,scale=0.5,baseline=0pt,xscale=0.9]
    \draw(0,0) node[tre] (r) {}; 
    \draw(-2,-1) node[tre] (c) {};
    \draw(-3,-2) node[tre] (b) {};
    \draw(-4,-3) node[tre] (a) {};
    \draw(2,-1) node[tre] (e) {};
    \draw(3,-2) node[tre] (f) {};
    \draw(4,-3) node[tre] (g) {};
    \draw(-3,-4) node[hyb] (A) {};
    \draw(-1,-4) node[hyb] (B) {};
    \draw(1,-4) node[hyb] (C) {};
    \draw(3,-4) node[hyb] (D) {};
    \draw(-3,-5) node[tre] (1) {};\etq 1
    \draw(-1,-5) node[tre] (2) {};\etq 2
    \draw(1,-5) node[tre] (3) {};\etq 3
    \draw(3,-5) node[tre] (4) {};\etq 4
    \draw(-1,-0.5) node[tre] (d) {};
    \draw(0,-5) node[tre] (5) {};\etq 5
    \draw[->](r)--(d);
    \draw[->](r)--(e);
    \draw[->](d)--(c);
    \draw[->](d)--(5);
    \draw[->](c)--(b);
    \draw[->](c)--(D);
    \draw[->](e)--(A);
    \draw[->](e)--(f);
    \draw[->](b)--(a);
    \draw[->](b)--(C);
    \draw[->](f)--(g);
    \draw[->](f)--(B);
    \draw[->](a)--(A);
    \draw[->](a)--(B);
    \draw[->](g)--(C);
    \draw[->](g)--(D);
    \draw[->](A)--(1);
    \draw[->](B)--(2);
    \draw[->](C)--(3);
    \draw[->](D)--(4);
  \end{tikzpicture}\quad
  \begin{tikzpicture}[thick,>=stealth,scale=0.5,baseline=0pt,xscale=0.9]
    \draw(0,0) node[tre] (r) {}; 
    \draw(-2,-1) node[tre] (c) {};
    \draw(-3,-2) node[tre] (b) {};
    \draw(-4,-3) node[tre] (a) {};
    \draw(2,-1) node[tre] (e) {};
    \draw(3,-2) node[tre] (f) {};
    \draw(4,-3) node[tre] (g) {};
    \draw(-3,-4) node[hyb] (A) {};
    \draw(-1,-4) node[hyb] (B) {};
    \draw(1,-4) node[hyb] (C) {};
    \draw(3,-4) node[hyb] (D) {};
    \draw(-3,-5) node[tre] (1) {};\etq 1
    \draw(-1,-5) node[tre] (2) {};\etq 2
    \draw(1,-5) node[tre] (3) {};\etq 3
    \draw(3,-5) node[tre] (4) {};\etq 4
    \draw(1,-0.5) node[tre] (d) {};
    \draw(0,-5) node[tre] (5) {};\etq 5
    \draw[->](r)--(d);
    \draw[->](r)--(c);
    \draw[->](d)--(e);
    \draw[->](d)--(5);
    \draw[->](c)--(b);
    \draw[->](c)--(D);
    \draw[->](e)--(A);
    \draw[->](e)--(f);
    \draw[->](b)--(a);
    \draw[->](b)--(C);
    \draw[->](f)--(g);
    \draw[->](f)--(B);
    \draw[->](a)--(A);
    \draw[->](a)--(B);
    \draw[->](g)--(C);
    \draw[->](g)--(D);
    \draw[->](A)--(1);
    \draw[->](B)--(2);
    \draw[->](C)--(3);
    \draw[->](D)--(4);
\end{tikzpicture}
\caption{\label{fig:non-time-cons}Non time consistent tree-sibling networks with the same $\mu$-representation.}
\end{figure}

Consider the phylogenetic networks depicted in
Fig.~\ref{fig:non-time-cons}. They are tree-sibling, binary, and the
single child of each hybrid node is a tree node; however, they do not
satisfy the time consistency condition. As it can be easily checked,
both networks have the same $\mu$-representation.

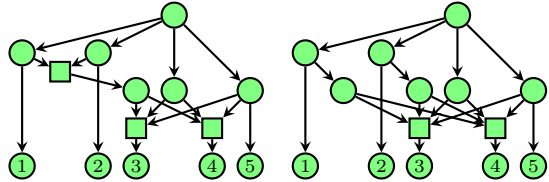
\begin{figure}
  \centering
  \begin{tikzpicture}[thick,>=stealth,scale=0.5,baseline=0pt]
    \draw(4,0) node[tre] (r) {};
    \draw(0,-1) node[tre] (a) {};
    \draw(2,-1) node[tre] (b) {};
    \draw(1,-1.5) node[hyb] (H) {};
    \draw(3,-2) node[tre] (c) {};
    \draw(4,-2) node[tre] (d) {};
    \draw(6,-2) node[tre] (e) {};
    \draw(3,-3) node[hyb] (A) {};
    \draw(5,-3) node[hyb] (B) {};
    \draw(0,-4) node[tre] (1) {}; \etq 1
    \draw(2,-4) node[tre] (2) {}; \etq 2
    \draw(3,-4) node[tre] (3) {}; \etq 3
    \draw(5,-4) node[tre] (4) {}; \etq 4
    \draw(6,-4) node[tre] (5) {}; \etq 5
    \draw[->](r)--(a);
    \draw[->](r)--(b);
    \draw[->](r)--(d);
    \draw[->](r)--(e);
    \draw[->](a)--(1);
    \draw[->](a)--(H);
    \draw[->](b)--(2);
    \draw[->](b)--(H);
    \draw[->](H)--(c);
    \draw[->](c)--(A);
    \draw[->](c)--(B);
    \draw[->](d)--(A);
    \draw[->](d)--(B);
    \draw[->](e)--(A);
    \draw[->](e)--(B);
    \draw[->](e)--(5);
    \draw[->](A)--(3);
    \draw[->](B)--(4);
  \end{tikzpicture}\quad
  \begin{tikzpicture}[thick,>=stealth,scale=0.5,baseline=0pt]
    \draw(4,0) node[tre] (r) {};
    \draw(0,-1) node[tre] (a) {};
    \draw(2,-1) node[tre] (b) {};
    \draw(1,-2) node[tre] (H) {};
    \draw(3,-2) node[tre] (c) {};
    \draw(4,-2) node[tre] (d) {};
    \draw(6,-2) node[tre] (e) {};
    \draw(3,-3) node[hyb] (A) {};
    \draw(5,-3) node[hyb] (B) {};
    \draw(0,-4) node[tre] (1) {}; \etq 1
    \draw(2,-4) node[tre] (2) {}; \etq 2
    \draw(3,-4) node[tre] (3) {}; \etq 3
    \draw(5,-4) node[tre] (4) {}; \etq 4
    \draw(6,-4) node[tre] (5) {}; \etq 5
    \draw[->](r)--(a);
    \draw[->](r)--(b);
    \draw[->](r)--(d);
    \draw[->](r)--(e);
    \draw[->](a)--(1);
    \draw[->](a)--(H);
    \draw[->](b)--(2);
    \draw[->](b)--(c);
    \draw[->](H)--(A);
    \draw[->](H)--(B);
    \draw[->](c)--(A);
    \draw[->](c)--(B);
    \draw[->](d)--(A);
    \draw[->](d)--(B);
    \draw[->](e)--(A);
    \draw[->](e)--(B);
    \draw[->](e)--(5);
    \draw[->](A)--(3);
    \draw[->](B)--(4);
  \end{tikzpicture}
  \caption{\label{fig:non-sb-TSTC}Non semi-binary, tree sibling, time consistent networks with the same $\mu$-representation.}
\end{figure}

Also the semi-binarity is a necessary condition, since
first
the network in
Fig.~\ref{fig:non-sb-TSTC} is time consistent and
tree-sibling, but not semi-binary, and has the same
$\mu$-representation as the second one, which is a sbTSTC network.

\begin{figure}
  \centering
  \begin{tikzpicture}[thick,>=stealth,scale=0.5,baseline=0pt]
    \draw(0,0) node[tre] (r) {};
    \draw(-2,-2) node[tre] (a) {};
    \draw(1,-1.6) node[tre] (b) {};
    \draw(2,-2) node[tre] (c) {};
    \draw(-1,-2) node[hyb] (A) {};
    \draw(0,-2) node[hyb] (B) {};
    \draw(-2,-4) node[tre] (1) {}; \etq 1
    \draw(-1,-4) node[tre] (2) {}; \etq 2
    \draw(1,-4) node[tre] (3) {}; \etq 3
    \draw(2,-4) node[tre] (4) {}; \etq 4
    \draw[->] (r)--(a);
    \draw[->] (r)--(b);
    \draw[->] (r)--(c);
    \draw[->] (a)--(A);
    \draw[->] (a)--(1);
    \draw[->] (A)--(2);
    \draw[->] (B)--(A);
    \draw[->] (c)--(B);
    \draw[->] (b)--(B);
    \draw[->] (b)--(3);
    \draw[->] (c)--(4);
  \end{tikzpicture}\quad
  \begin{tikzpicture}[thick,>=stealth,scale=0.5,baseline=0pt]
    \draw(0,0) node[tre] (r) {};
    \draw(-2,-2) node[tre] (a) {};
    \draw(1,-1.6) node[tre] (b) {};
    \draw(2,-2) node[tre] (c) {};
    \draw(-1,-2) node[hyb] (A) {};
    \draw(0,-2) node[hyb] (B) {};
    \draw(-2,-4) node[tre] (1) {}; \etq 1
    \draw(-1,-4) node[tre] (2) {}; \etq 2
    \draw(1,-4) node[tre] (3) {}; \etq 3
    \draw(2,-4) node[tre] (4) {}; \etq 4
    \draw[->] (r)--(a);
    \draw[->] (r)--(b);
    \draw[->] (r)--(c);
    \draw[->] (a)--(A);
    \draw[->] (a)--(3);
    \draw[->] (A)--(2);
    \draw[->] (B)--(A);
    \draw[->] (c)--(B);
    \draw[->] (b)--(B);
    \draw[->] (b)--(1);
    \draw[->] (c)--(4);
  \end{tikzpicture}
  \caption{\label{fig:hyb-hyb}Networks with hybrid children of hybrid nodes and the same
    $\mu$-representation.}
\end{figure}
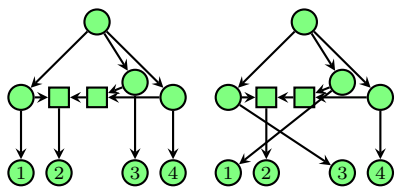

To conclude with this series of counterexamples, the condition that
the single child of a hybrid node is a tree node is also necessary, as
the networks in Fig.~\ref{fig:hyb-hyb}, both with the same
$\mu$-representation, show. 

\section{Conclusions}
\label{sec:concl}


While there exist in the literature some algorithms to
reconstruct sbTSTC phylogenetic networks
from biological
sequences, no distance metric was known in this class that is both
mathematically consistent and computationally efficient.
The $\mu$-distance we have defined fulfills these two
requirements, and is already implemented in a package included in the
BioPerl bundle. 

This $\mu$-distance is based on the $\mu$-representation of networks:
a multiset of vectors of natural numbers, each of them associated
to a node.
This $\mu$-representation could also be used to define
alignments between phylogenetic networks
\cite[Sec.~VI]{cardona.ea:07b}, which are useful in order to display
at a glance the differences between alternative evolutionary histories
of a set of species. Some results in this direction will be shortly
published elsewhere.

As a by-product, we have also obtained a procedure to generate all the
sbTSTC networks on a given set of taxa up to isomorphism. We are
working in an efficient implementation for their generation, in order to
include it in a forthcoming release of BioPerl.


\end{document}